\documentclass{article}
\usepackage{graphicx} % Required for finserting images

\usepackage{caption}
\usepackage{subcaption}
\usepackage{amsmath}
\usepackage{color,colortbl}

\usepackage[english]{babel}
\usepackage{csquotes}
\usepackage{comment}

\usepackage[top=2cm,bottom=2cm,left=0.98in,right=0.79in,marginparwidth=1.75cm]{geometry}
\usepackage{amsmath}
\usepackage[backend = biber,
    style = numeric, %numeric, alphabetic
    giveninits = true,
    natbib = true,
    hyperref = true,
    maxbibnames = 11, % number of authors shown
    ]{biblatex} % (taken from https://github.com/hoelzer/dfg)

\usepackage[parfill]{parskip}
\usepackage[table,usenames,dvipsnames,svgnames]{xcolor}
\definecolor{darkmidnightblue}{rgb}{0.0, 0.2, 0.4}
\usepackage[colorlinks=true, linkcolor=MidnightBlue, citecolor=MidnightBlue, urlcolor=MidnightBlue, pdfpagelayout=SinglePage, bookmarks, bookmarksopen]{hyperref}
\newcommand{\LTL}{Learning-to-learn}

\newcommand{\Ca}{$\text{Ca}^{2+}$}
\usepackage{float}
\usepackage{authblk}

\usepackage{soul}
\definecolor{elenaorange}{rgb}{1,0.5,0}

\definecolor{pierblue}{rgb}{0.0,0.0,1.0}
\newcommand{\pier}[1]{{\color{pierblue}{#1}}}
\definecolor{wwgreen}{rgb}{0.1,0.6,0.0}

\title{Two-compartment neuronal spiking model expressing brain-state specific apical-amplification, -isolation and -drive regimes}

\author [1] {Elena Pastorelli}
\author [2,5] {Alper Yegenoglu}
\author [1,3] {Nicole Kolodziej}
\author [4] {Willem Wybo}
\author [1] {\\Francesco Simula}
\author [2] {Sandra Diaz}
\author [6] {Johan Frederik Storm}
\author [1] {Pier Stanislao Paolucci}

\affil[1]{Istituto Nazionale di Fisica Nucleare, Sezione di Roma, Roma, Italy}
\affil[2]{Simulation and Data Lab Neuroscience, Jülich Supercomputing Centre (JSC), Institute for Advanced Simulation, JARA, Jülich Research Center, Jülich, Germany}
\affil[3]{Dipartimento di Fisica, Università di Roma Sapienza, Roma, Italy}
\affil[4]{Institute of Neuroscience and Medicine (INM-6) and Institute for Advanced Simulation (IAS-6) and JARA-Institute Brain Structure–Function Relationships (INM-10), Jülich Research Center, Jülich, Germany}
\affil[5]{Department of Mathematics, Institute of Geometry and Applied Mathematics, RWTH Aachen University, Aachen, Germany}
\affil[6]{Department of Molecular Medicine, Institute of Basic Medical Sciences, University of Oslo, Oslo, Norway}

\begin{document}

\maketitle 
\begin{abstract}
Mounting experimental evidence suggests that brain-state-specific neural mechanisms, supported by connectomic architectures, play a crucial role in integrating past and contextual knowledge with the current, incoming flow of evidence (e.g., from sensory systems). These mechanisms operate across multiple spatial and temporal scales, necessitating dedicated support at the levels of individual neurons and synapses. A notable feature within the neocortex is the structure of large, deep pyramidal neurons, which exhibit a distinctive separation between an apical dendritic compartment and a basal dendritic/perisomatic compartment. This separation is characterized by distinct patterns of incoming connections and brain-state-specific activation mechanisms, namely, apical amplification, isolation, and drive, which are associated with wakefulness, deeper NREM sleep stages, and REM sleep, respectively. The cognitive roles of apical mechanisms have been demonstrated in behaving animals. In contrast, classical models of learning in spiking networks are based on single-compartment neurons, lacking the ability to describe the integration of apical and basal/somatic information. This work aims to provide the computational community with a two-compartment spiking neuron model that incorporates features essential for supporting brain-state-specific learning. This model includes a piece-wise linear transfer function (ThetaPlanes) at the highest abstraction level, making it suitable for use in large-scale bio-inspired artificial intelligence systems. A machine learning evolutionary algorithm, guided by a set of fitness functions, selected the parameters that define neurons expressing the desired apical mechanisms.
\end{abstract}
\section{Introduction}

%Association of internal prediction and perception at system and cellular level (PSP, JS)
Thanks to an evolutionary history spanning hundreds of millions of years and selecting from countless individuals, the structural connectome and cellular mechanisms have become adept at supporting the integration of multi-modal sensory evidence with internal hypotheses about the world and the self \cite{PapaleMuckli2023, MuckliPetroGoebel2015, SpornsTononiKötter2005, HagmannCammounSporns2008}. Additionally, specialized solutions have emerged at the macro-, meso-, and micro-scales, enabling the expression of dynamic repertoires of functional connectivity \cite{CabralDeco2017}. At the cellular level, within large, cortical pyramidal cells of the mammalian neo-cortex, specific feed-forward sensory input is combined with contextual and feed-back information by the \textit{apical-amplification} principle \cite{2015PhillipsLarkumSilverstein, larkum2013, larkum1999, Larkum2009}, with \cite{phillips2023} assuming apical mechanisms to be among key cellular foundations of the mental life. While this type of amplification seems to dominate during wakefulness \cite{larkum2013}, evidence suggests that it is replaced by different principles and mechanisms during transitions to other brain-states \cite{aruSuzukiLarkum2020, aruSiclariStorm2020}, namely \textit{apical-isolation} during the deepest stages of NREM sleep (like in anesthesia \cite{SuzukiLarkum2020}) and \textit{apical-drive} during dreaming \cite{aruSuzukiLarkum2020}).

%Biophysical mechanisms supporting brain state specific regimes  (JS, PSP)
The existence of different brain states, supported by state-specific cellular and systemic mechanisms, is also of ancient origin. Sleep has withstood the evolutionary pressure across all studied animal species, despite its apparent lack of productivity. It promotes memory consolidation and integration, as well as preparation for anticipated tasks \cite{tononiCirelli2014,Buzsaki2015,SejnowskiDestexhe2000}, and returns the network to optimal working points after periods of awake learning \cite{WatsonButzsaki2016,TononiCirelli2019}. Mammals devote a significant portion of their time to sleep, especially youngsters who learn at the fastest rate \cite{FrankIssaStryker2001}. Moreover, sleep deprivation negatively impacts cognitive performance \cite{Killgore2010}. These considerations underscore the importance of detailed modeling of sleep's cognitive functions and the underlying cellular mechanisms.

%Limits of single compartment in spiking networks engaged in learning and sleep cycles (PSP)
Here, we propose a method to transition from the classical modeling approach of networks, which relies on single-compartment neurons, towards incorporating simple apical \Ca-dynamics. This inclusion supports the expression of intriguing brain-state-specific learning capabilities. Single-compartment models with spike frequency adaptation, such as the Adaptive Exponential Integrate and Fire neuron (AdEx) \cite{BretteGerstnerAeif_cond_alpha2005}, have enabled the construction of networks capable of entering both wakefulness-like asynchronous irregular regimes and deep-sleep-like synchronous slow oscillation regimes (e.g., \cite{Pastorelli2019}, \cite{CaponeRebolloSanchezVivesMattia2019}). For such networks, mean-field models have been developed \cite{DiVoloRomagnoniCaponeDestexhe2019}. These mean-field descriptions of the behavior of spiking networks composed of AdEx neurons have supported the development of models based on connectomes at the scale of the whole brain \cite{DiVoloRomagnoniCaponeDestexhe2019,AquilueGoldmanDestexhe2023}, also capable of expressing both the asynchronous and synchronous regimes. However, these models do not capture the activity of individual neurons and synapses in engram coding, nor do they support the simulation of the temporal evolution of engrams \cite{JosselynFranklan2015}.

The cognitive and energetic functions specific to different brain states have been explored in spiking models engaged in learning and sleep cycles. These models aim to simulate the activity and contribution of individual neurons and monitor synaptic changes over time \cite{caponePastorelliGolosioPaolucci2019, golosioDelucaCaponePastorelliPaolucci2021, delucaTonielliPastorelliCaponePaolucci2023}. Although these models utilize the temporal coincidence between contextual and perceptual information, they are still based on on single-compartment neurons. Therefore, they necessitate precise calibration of currents carrying contextual priors and novel evidence. Such modeling approaches cannot fully leverage the capabilities of apical mechanisms, for example, the transition to much higher frequencies associated with \textit{apical-amplification} during wakefulness, \textit{apical-drive} during dreaming, or \textit{apical-isolation} during deep, slow-wave sleep.

%link to bio-inspired AI
Within the framework of bio-inspired artificial intelligence, a few studies (e.g., \cite{CaponeLupoMuratorePaolucci2023,CaponeLupoMuratorePaolucci2022}) have begun to explore the specific advantages of apical-amplification-like bursting mechanisms for fast learning in spiking networks engaged in complex temporal tasks. However, these models have taken as working hypotheses the existence of transfer functions that enter a bursting regime when a temporal coincidence between perceptual and sensorial signals is detected. Here, we demonstrate how to construct a two-compartment neuron based on cellular biophysical evidence, capable of supporting the apical-amplification bursting mechanism.
% A simplified transfer function is useful for AI application and mean-field models
Furthermore, bio-inspired Artificial Intelligence (AI) algorithms would benefit from neural models characterized by a simple transfer function, simplifying the definition of training rules. A classic transfer function adopted in AI algorithms is the ReLU (rectified linear unit) rule, which approximates the transfer function of single-compartment neurons.
% ThetaPlanes: a generalization of the ReLU
We will show how to introduce a transfer function suitable for approximating the response of the two-compartment neuron to the combination $(I_s, I_d)$ of somatic and distal signals, capable of describing the apical-amplification, -isolation, and -drive regimes. We have named this transfer function \textit{ThetaPlanes$(I_s, I_d)$}.  

%Searching in high dimensionality space (SD, EP)
The extension of the AdEx model to include an apical compartment with simplified \Ca-dynamics (the Ca-hotzone, here abbreviated to Ca-HZ) requires a few tens of parameters, implying a search in a high-dimensional space for fine-tuning. For any mathematical model, understanding the sensitivity of the model output to perturbations and correlations among the parameters defining it is crucial. This need becomes even more apparent when dealing with high-dimensional parameter spaces, where the dependency of outputs on underlying parameters becomes less intuitive for the modeler. As in many other research fields, neuroscience demands a thorough understanding of these relationships to draw meaningful conclusions about the simulated behavior of the modeled phenomena \citep{nowke2018toward, Yegenoglu2022exploring}. Population-based optimization techniques offer a more efficient approach to exploring large parameter spaces than brute-force testing of all possible parameter combinations. Depending on the shape of the manifolds, different algorithms may be more or less effective in navigating the parameter space and identifying areas of interest to the modeler. While, for example, gradient-based methods typically identify local minima and converge very quickly, not all fitness evaluation measures and parameter spaces are suitable for such algorithms \citep{yegenoglu2020ensemble}. Simulated annealing and cross-entropy methods provide suitable gradient-free exploration techniques but also require fine-tuning of hyperparameters. Evolutionary strategies and similar population-based methods can effectively navigate complex parameter spaces and quickly adapt to the manifolds if the level of noise or stochasticity is maintained at a suitable level, depending on the variations induced by the parameters with respect to the fitness. Several such algorithms can be tested and even combined to achieve a comprehensive understanding of parameter sensitivity and interdependencies. The tools and methodology adopted in this work to explore the parameter space defining the two-compartment model we named \textit{Ca-AdEx}, and the evolutionary approach based on the definition of a \textit{genome} and a \textit{fitness function}, are detailed in dedicated subsections of \nameref{sec:Methods}.

%Multi-compartment neuronal models (WW)
Multi-compartment (MC) models have been successful in reproducing experimentally observed dendritic processes and computations \cite{Segev2000}, particularly the interaction between apical \Ca-spikes and somatic action potentials \cite{Hay2011}. Most often, MC models are paired with Hodgkin-Huxley (HH) type ion channels. The spatially extended nature of the dendritic tree, requiring many compartments, leads to models that are expensive to simulate. Past simplification efforts have focused on two largely orthogonal axes of advance: either condensing the HH channels into a simpler effective spike generation mechanism \cite{Kistler1997, Pozzorini2015} or reducing the number of compartments needed in a simulation while maintaining desired response properties \cite{Wybo2021}. To ultimately arrive at the most efficient formulation of a neuron model, a simplified description of dendritic non-linearities needs to be combined with a reduction in the number of compartments, in such a way that the model architecture is flexible and can admit a range of dendritic computations. Previous work on this topic used a hybrid combination of compartment dynamics and kernel convolutions \cite{Naud2014}, the former to model \Ca-activation and the latter to capture the somato-dendritic interactions. While the use of convolutions is a general way to capture the linear component of intra-dendritic interactions \cite{Wybo2013b, Wybo2015}, it is computationally inefficient compared to the use of normal coupling terms between compartments \cite{Wybo2021}. For this reason, we propose an approach that solely relies on normal compartmental dynamics, which has the added advantage of potentially integrating any type of nonlinear conductance. By design, this approach can thus also implement other dendritic non-linearities, such as N-Methyl-D-Aspartate (NMDA) spikes \cite{Schiller2000, Major2008, Major2013}. We demonstrate this potential by extending the two-compartment Ca-AdEx model to a multi-compartment description, which, next to the Ca-HZ and soma compartments, features apical and basal compartments suited for NMDA-spike generation. Furthermore, we have implemented a compartmental modeling framework in NEST \citep{Gewaltig2007, Spreizer2022} that supports the aforementioned \Ca-, AdEx-, and NMDA dynamics. Combined, our work facilitates the study of dendritic dynamics with simplified neuron models at the network level.

\section{Methods}
\label{sec:Methods}

\subsection{The two-compartment Ca-AdEx model supporting calcium spike firing}
\label{subsec:TwoCompModel}
One of the focal points of this endeavor was the creation of a neuron model able to express properties of apical amplification during awake states, to aid the formation of memories inside the synaptic matrix during incremental learning cycles. Indeed, recent studies (\cite{aruSiclariStorm2020,aruSuzukiLarkum2020}) have highlighted the critical role of apical amplification for conscious processing during the awake state in layer 5 pyramidal neurons (L5PC), in contrast with the mechanisms of apical drive and apical isolation that are predominant respectively in REM and NREM sleep.
To replicate these states, it is essential to have an apical compartment able to support \Ca spike, (\cite{larkum1999,larkum2013}), considered as the cellular mechanism underpinning apical amplification. Meanwhile, the soma follows the dynamics of an adaptive exponential integrate and fire neuron (AdEx), described by the following equations (\cite{gerstner_kistler_naud_paninski_2014}):

\begin{equation}
\label{eq:AdEx}
\left\{
\begin{array}{ll}
    C_m \frac{dV}{dt} & =
 -g_L(V-E_L)+g_L\Delta_T\exp\left(\frac{V-V_{th}}{\Delta_T}\right) -
 g_e(t)(V-E_e)-g_i(t)(V-E_i)-w +I_e 
 \\
 \\
 \tau_w \frac{dw}{dt} & = a(V-E_L) +b\sum_{k}\delta (t-t_{k}) - w
 \end{array}
\right.
\end{equation} 
The parameters are detailed in the \textit{Soma passive parameters} section of table \ref{tab:GenomeTable}, while $I_e$ represents all the external currents.

%Support for calcium spike in distal compartment
%What is it BAC firing
The backpropagation-activated calcium spikes (BAC firing) is induced by the coincidental occurrence of a synaptic input to the apical dendrite and a spike generated within the soma. This spike backpropagates to the Ca-HZ within the apical dendrite (BAP), effectively lowering the threshold required for a dendritic \Ca-spike. Consequently, this mechanism can trigger a burst of multiple action potentials, even in the presence of a subthreshold distal excitatory postsynaptic signal.

%BAC firing can be ideally seen as the associations of feed-forward and feedback cortical signals

%What is needed
The activation of the calcium spike in the dendrite is the critical element for the BAC firing. To support this activation, we modeled a neuron implementing a voltage dependent \Ca current and the \Ca concentration dynamics within the apical dendritic compartment (Ca-HZ). Additionally, along a \Ca-activated $K$ current is included to re-polarize the dendritic membrane and terminate the \Ca-spike. 
%In addition, the activation of the BAC firing requires the implementation of a back-propagating action potential that is able to reach the dendritic zone of \Ca activation, allowing the evocation of the calcium spike.

%Intracellular [Ca] dynamics
The dendritic intracellular \Ca concentration dynamics has been modeled, as described in \cite{gerstner_kistler_naud_paninski_2014}, using the following equation:

\begin{equation}
    \label{eq:calcium_conc_Larkum}
    \frac{d[Ca]}{dt} = \phi_{Ca}I_{Ca} + \frac{[Ca] - [Ca]_0}{\tau_{Ca}}
\end{equation}
where $[Ca]_0$ represents the baseline of the intracellular \Ca concentration in mM, $\tau_{Ca}$ is the time constant of calcium extrusion in ms, $I_{Ca}$ is the high voltage activated \Ca current circulating in the dendrite in pA and $\phi_{Ca}$ is a scaling factor.

%High voltage activated Ca current
The dendritic ion currents were modeled using the Hodgkin-Huxley formalism. The high voltage activated \Ca current ($I_{Ca}$) has been modeled as in \cite{larkum2004}:
\begin{equation}
\label{eq:Ca_curr}
I_{Ca} = g_{Ca}mh(E_{Ca}-V)
\end{equation}
where $g_{Ca}$ is the maximal calcium conductance in nS, $E_{Ca}$ is the calcium reversal potential and $V$ the membrane voltage, both in mV.  The activation and inactivation variables, $m$ and $h$ respectively, are characterized by first-order kinetics:
\begin{equation}
\label{eq:Ca_mh_variables}
\frac{dm}{dt} = \frac{m_{\infty} - m}{\tau_{m}} 
\text{ and }
\frac{dh}{dt} = \frac{h_{\infty} - h}{\tau_{h}}
\end{equation}
where $m_{\infty}$ and $h_{\infty}$ are the corresponding steady state functions and $\tau_{m}$ and $\tau_{h}$ are their time constants in ms. The steady state functions are given by:
\begin{equation}
\label{eq:Ca_mh_inf}
m_{\infty} = \frac{1}{1 + exp(m_{slope}(V - (m_{half})))}
\text{ and }
h_{\infty} = \frac{1}{1 + exp(h_{slope}(V - (h_{half})))}
\end{equation}
with $m_{slope}$ and $h_{slope}$ representing the slope of the two functions and $m_{half}$ and $h_{half}$ representing the half activation/deactivation values in mV.

%Ca activated K current
The \Ca activated $K$ current ($I_{K_{Ca}}$) has been modeled as in \cite{Hay2011}:
\begin{equation}
\label{eq:K_curr}
I_{K_{Ca}} = g_{K}m(E_{K}-V)
\end{equation}
where $g_{K}$ is the maximal potassium conductance in nS, $E_{K}$ is the potassium reversal potential and $V$ is the membrane voltage, both in mV. $m$ represents the activation variable described by the first order kinetics:
\begin{equation}
\label{eq:K_m_variable}
\frac{dm}{dt} = \frac{m_{\infty} - m}{\tau_{m}}
\end{equation}
Here $\tau_{m}$ is the potassium time constant in ms and $m_{\infty}$ is the activation steady state variable described by:
\begin{equation}
\label{eq:K_m_inf}
m_{\infty} = \frac{1}{1 + (\frac{Ca_{th}}{[Ca]})^{exp_{K_{Ca}}}}
\end{equation}
where $Ca_{th}$ represents the Ca concentration threshold for calcium channel opening in mM and $exp_{K_{Ca}}$ is an exponential factor.
% Add something about the possible values assumed by the exponential factor

In summary, a simple two-compartment Ca-AdEx neuron is described by equation (\ref{eq:Ca-AdEx}) in Section \ref{subsec:Ca-AdExEquations}.

The AdEx mechanism and \Ca currents were implemented within the NEST compartmental modelling framework (\ref{subsec:SupportInNEST}), allowing their incorporation in the somatic and Ca-HZ compartment, respectively.

%Desired transfer function
% relation with fitness function

\subsection{The \textit{genome} of the Ca-AdEx model}
\label{subsec:Genome}

The behaviour of a neuron model is characterized by a set of parameters required to describe its dynamics. In this paper we named this set of parameters as the \textit{genome} of the neuron, because it has been identified using an evolutionary algorithm.
For the multi-compartment neuron model, this genome comprises all parameters, both passive and active, necessary to define the dynamics of each compartment and all the ionic currents involved. 

The passive parameters which pertain to membrane properties, allow neurons to conduct electrical impulses without the use of voltage-gated ion channels. These parameters detail the membrane potential changes in response to currents across the cell membrane.
Among the passive neuron parameters, we include the capacitances ($C_m$), leak conductances ($g_L$, i.e. conductances that do not vary with the membrane potential or other parameters), and reversal potentials of both somatic and dendritic compartments ($E_L$). These are responsible for the under-threshold and spike-triggered dynamics described in equation \ref{eq:AdEx}.

The parameters governing the ionic currents are defined as active parameters. In the model utilized in this work, the active parameters encompass all those used to describe the dynamics of calcium concentration, the voltage dependent calcium current and the calcium activated potassium current.

Table \ref{tab:GenomeTable} presents the complete genome used to describe the Ca-AdEx neuron defined in this work. 

Section \ref{sec:ExemplaryParams} reports the genome of the best Ca-AdEx neuron identified by the evolutionary search, used in next sections, if not otherwise stated.  

\begin{table}[H]
\caption{Neuron Genome: parameters characterizing the Ca-AdEx neuron. See Section \ref{sec:ExemplaryParams} for the values of the neuron identified by the evolutionary search.}
\label{tab:GenomeTable}
\begin{center}
\begin{tabular}[t]{|c|c|c|}
\hline
\multicolumn{3}{|c|}{\textbf{Soma passive parameters}}\\ 
\hline
$C_m^s$ & Membrane capacitance & pF\\
\hline
$g_L^s$ & Leakage conductance & nS\\
\hline
$E_L^s$ & Leakage reversal potential & mV\\
\hline
$t_{ref}$ & Refractory period & ms\\
\hline
$\Delta_T$ & Slope factor & mV\\
\hline
$a$ & Subthreshold adaptation & nS\\
\hline
$b$ & Spike-triggered adaptation & pA\\
\hline
$\tau_w$ & Adaptation time constant & ms\\
\hline
$V_{th}$ & Membrane voltage threshold & mV\\
\hline
$V_{reset}$ & Membrane voltage after-spike reset & mV\\
\hline
$w_{BAP}$ & BAP amplitude & mV\\
\hline
$d_{BAP}$ & BAP delay & ms\\
\hline
\multicolumn{3}{|c|}{\textbf{Distal passive parameters}}\\ 
\hline
$C_m^d$ & Membrane capacitance & pF\\
\hline
$g_L^d$ & Leakage conductance & nS\\
\hline
$g_C$ & Soma-distal coupling conductance & nS\\
\hline
$E_L^d$ & Resting potential & mV\\
\hline
\multicolumn{3}{|c|}{\textbf{Distal active parameters}}\\ 
\hline
$\Bar{g}_{Ca}$ & Max Ca conductance & nS\\
\hline
$\tau_{Ca}$ & Ca decay time constant & ms\\
\hline
$\tau_{m}$ & Ca activating function time constant & ms\\
\hline
$\tau_{h}$ & Ca deactivating function time constant & ms\\
\hline
$m_{half}$ & Ca activating function half voltage & mV\\
\hline
$h_{half}$ & Ca deactivating function half voltage & mV\\
\hline
$m_{slope}$ & Ca activating function slope & -\\
\hline
$h_{slope}$ & Ca deactivating function slope & -\\
\hline
$[Ca]_{th}$ & Ca concentration threshold for Ca channel opening & mM\\
\hline
$[Ca]_0$ & Baseline intra-cellular Ca concentration & mM\\
\hline
$\phi$ & Scaling factor in Ca concentration dynamics & -\\
\hline
$\bar{g}_{K_{Ca}}$ & Maximal conductance of Ca dependent K current & nS\\
\hline
$\tau_{K_{Ca}}$ & Activating function time constant of Ca dependent K current & ms\\
\hline
$exp_{K_{Ca}}$ & Exponential factor in Ca dependent K current & -\\
\hline
$E_K$ & K reversal potential & mV\\
\hline
\end{tabular}
\end{center}
\end{table}

% NOTE: here gbar_Na_AdEx is not considered an independent parameter but it is put equal to g_L_s

\subsection{Fitness functions}
\label{subsec:FitnessFunc}
Specific fitness functions have been devised to constrain the model. These functions aim to guide the evolutionary search within parameter space toward optimal configurations, focusing on the identification of neurons that embody both the spiking frequency–stimuli relationships characteristic of apical mechanisms and a response to somatic-only stimulus that mirrors the behaviour of a single-compartment AdEx neuron.

In our work, the model's fitness was assessed considering two different tasks: response to \textit{pulse stimuli} of a few milliseconds in duration and response to \textit{prolonged stimuli} lasting few seconds. 

In the \textit{pulse stimuli} task, the Ca-AdEx neuron model is designed to replicate the experiment proposed in \cite{larkum1999}, demonstrating the apical amplification effects through the activation of the BAC firing in response to short-duration currents. The goal is to emulate the observations reported in the four panels of figure 1 from the cited study. These panels depict the response to four combination of short duration inputs delivered to the apical and somatic compartment. Notably, the most interesting scenario involves injecting a threshold step current into the somatic compartment for $5ms$, accompanied by an under-threshold \textit{beta} current in the distal compartment with a $5ms$ delay. The threshold somatic current's amplitude is calibrated elicit a single spike in isolation. Conversely, the provision of the under-threshold distal current alone does not produce any spike. The essential behaviour to replicate is that the combination of these two currents can activate the BAC firing mechanism, leading to a high-frequency burst of three spikes (see Figure \ref{fig:pulseProtocol}). To guide the model towards accurately responding to the four combination of short-duration pulses, four fitness functions are employed. These functions aim to generate the correct number of spikes in short-duration bursts and to delineare a regime of under-threshold distal stimulus (see the \textit{Pulse stimuli} section of table \ref{tab:FitnessTable}). 

In the second optimization task, the \textit{prolonged stimuli} task, the neuron model is subjected to pairs of prolonged-duration $(I_s, I_d)$ DC input currents. Combinations of somatic and distal stimuli are kept constant for $2s$, followed by a $3s$ period of zero input. The corresponding set of fitness functions is detailed in the \textit{Prolonged stimuli} section of table \ref{tab:FitnessTable}. In this scenario, the computation of fitness functions relies on several different measures. Initially, evaluations are made concerning the activation of \Ca channels and, following activation, their closure after the stimulus concludes. Moreover, individuals (i.e., model configurations) that activate calcium spikes even with purely somatic currents are excluded by a dedicated fitness function.

Then, our goal is to develop a two-compartment neuron that, when stimulated somatically, mimics an equivalent single-compartment AdEx neuron. To achieve this, we defined two fitness functions. The first employs the Earth Mover's Distance (EMD) algorithm. Additionally, we compare the rheobase of the single-compartment and two-compartment neurons (refer to the \textit{AdEx matching} subsection of table \ref{tab:FitnessTable}).

An additional set of fitness functions aims to ensure the model exhibits a high, linear gain associated with the apical mechanism (refer to \textit{Gain \& linearity of apical mechanism} in table \ref{tab:FitnessTable}). Specifically, for the $\nu(I_s,I_d)$ transfer function, evaluations include: the firing rate following \Ca opening for a distal-only stimulus $(I_s=0,I_d)$, and the linearity in the firing rate increase linked to calcium channel activation for increasing somatic and distal currents. Moreover, particular fitness functions focus on ensuring the monotonicity of the $\nu(I_s,I_d=const)$ curves and the presence of the apical gain mechanism across the desired input domain: $I_s={0,..,I_s^{Max}}$, $I_d={0,..,I_d^{Max}}$.

A fitness function is dedicated to excluding neurons exhibiting "epileptic" behavior when stimulated within the predefined range of currents (refer to the \textit{Exclusion of pathological configurations} section in table \ref{tab:FitnessTable}). Furthermore, an additional set of \textit{Cautionary checks} is introduced to further constrain the neurons. However, these additional constraints have likely been redundant in the context of our numerical experiments.   

\begin{table}[H]
\caption{Fitness functions}
\label{tab:FitnessTable}
\begin{center}
\begin{tabular}[t]{|c|c|}
\hline
\multicolumn{2}{|c|}{\textbf{~~~~~~~~~~Pulse stimuli task}}\\
\hline
$L2\_PD$ & Number of spikes for (th $I_s, I_d$ under-th) [target=3]\\
\hline
$L2\_PE$ & Number of spikes for ($I_s=0,I_d$ over-th) [target=2]\\
\hline
$L2\_PG$ & Diff. of spikes between ($I_s, I_d$ under-th) and ($I_s=0,I_d$ over-th) [target=1]\\
\hline
$L2\_PR$ & $I_d$ over-threshold / $I_d$ under-threshold ratio\\
\hline
\multicolumn{2}{|c|}{\textbf{~~~~~~~~~~Prolonged stimuli task}}\\ 
\hline
  & \textit{Primary checks of apical channels activation}\\
\hline
$L2\_CaO$ & Missed Ca channel opening \\
\hline
$L2\_CaC$ & Missed closure of apical mechanism \\
\hline
$L2\_CaO\_soma$ & Check for Ca NOT opening for $I_d = 0$ \\
\hline
  & \textit{AdEx matching}\\
\hline
$L2\_SEMD$ & EMD between adex and soma curves, target $=0$ \\
\hline
$L2\_SH$ & Somatic rheobase, target is single-compartment AdEx threshold \\
\hline
  & \textit{Gain \& linearity of apical mechanism}\\
\hline
$L2\_120CaH$ & Distal firing rate after Ca opening \\
\hline
$L2\_NUMJUMPS$ & Number of jumps in f/I curves \\
\hline
$L2\_LinJump$ & Linearity in Ca jump for growing (Is,Id) currents \\
\hline
$L2\_MONOTON$ & Monotonicity of the f/I curves \\
\hline
  & \textit{Exclusion of pathological configurations}\\
\hline
$L2\_E$ & Check for epileptic neuron \\
\hline
  & \textit{Cautionary checks}\\
\hline
$L2\_MINV$ & Minimum voltage, target is adex minimum voltage \\
\hline
$L2\_CSG\_R$ & Gain in in f/I curves due to calcium spike \\
\hline
$L2\_DS\_R$ & Ratio between somatic and distal thresholds, when distal firing rate > 4Hz \\
\hline
$L2\_SHZD$ & Ratio between somatic and distal thresholds \\
\hline
\end{tabular}
\end{center}
\end{table}

\subsection{The Learning to Learn framework}
\label{subsec:L2L}

\begin{figure}[ht!]
  \centering
  \includegraphics[width=0.6\textwidth]{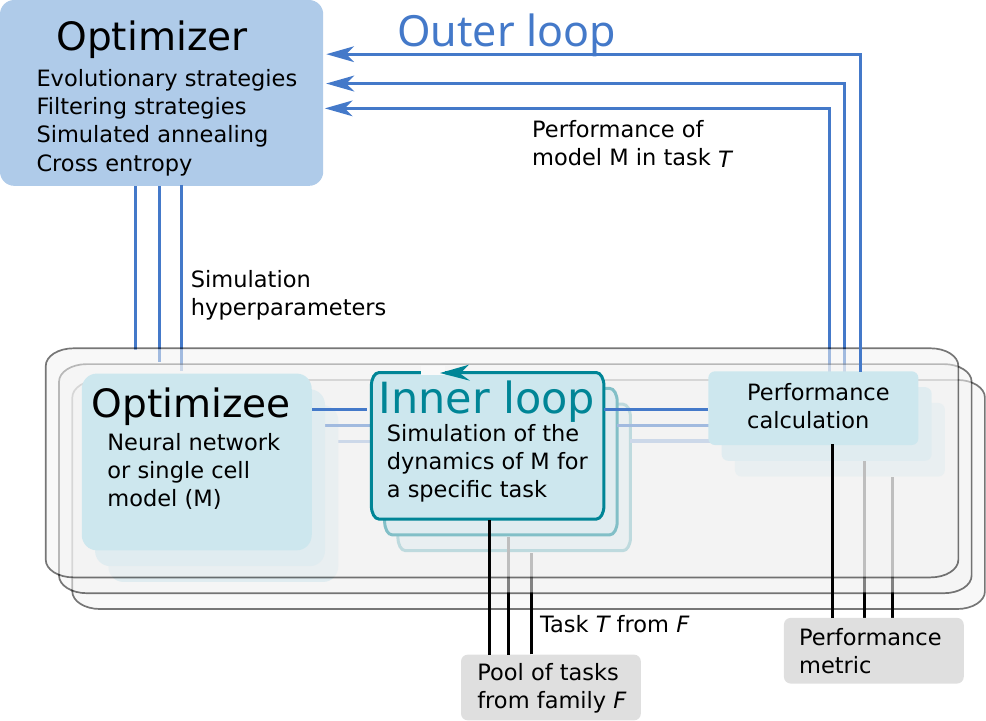}
  \caption{\label{fig:l2l_loops} The two-loop scheme of L2L. In the inner loop, a model is trained or simulated on a task from a family of tasks. A fitness function evaluates the performance of the model. The model parameters are optimized in the outer loop. Image provided by \cite{Yegenoglu2022exploring}.}
\end{figure}

\LTL, or meta-learning \parencite{thrun2012learning,thrun1998learning}, is an approach in machine learning aimed at enhancing learning performance through generalization. In a conventional learning setting, a program or algorithm is trained to perform a single task, evaluated by a specific performance metric. The algorithm's performance improves as it is exposed to more training samples. After sufficient training, the algorithm or model can achieve high performance on new samples of the same task that it did not encounter during the training phase.

In learning-to-learn, this paradigm is broadened into a two-loop structure, as shown in Figure~\ref{fig:l2l_loops}. In the inner loop, the program, also known as the \textit{optimizee}, can  adapt to learn a specific task from a family of tasks. These tasks may range from classification and inference to training multi-agents for complex problem-solving. A fitness function assesses the performance of the optimizee and yields a fitness value. This function is tailor-made for the task and must be precisely defined to effectively evaluate the optimizee. In the outer loop, the algorithm's overall performance is enhanced by optimizing the hyper-parameters or parameters across a spectrum of tasks, facilitating the evolution of the entire system.

In \cite{Yegenoglu2022exploring}, we introduced an implementation of the learning-to-learn concept within a framework named \textit{L2L}. In L2L, the outer loop is composed of various gradient-free optimization techniques based on metaheuristics, including evolutionary algorithms or filtering strategies. The framework's versatility allows for the execution of any algorithm or simulation, which can then be operated on anything from local machines to high-performance computing systems (HPCs). Thanks to the framework's inherently parallel structure, multiple instances of the inner loop can be efficiently deployed on HPC systems. L2L necessitates only a performance measure and a set of parameters for optimization targets. It is developed in Python, is available as open-source, and adheres to an open development model.

\subsection{Execution Environment of L2L on HPC Platforms}
\label{subsec:L2LExecEnv}

L2L is equipped to iteratively deploy instances of the inner loop on HPC resources in a variety of ways. It is compatible with any scheduler present on a cluster or supercomputer. For this project, deployment occurred on the JUSUF supercomputer at the Jülich Supercomputing Center, as well as on the local cluster at the University of Rome. In both instances, slurm served as the scheduler to allocate the necessary computational resources.

%Depending on the complexity of the parallelization required in the inner loop, there are different options for executing each individual. 

In this work, for each optimization run we request a single allocation which comprises enough computational resources to launch all individuals in each iteration of the outer loop. Subsequently, L2L launches $NS$ steps within the job allocation to distribute the resources among the individuals. This distribution is achieved by setting the appropriate scheduler parameters within the ``exec'' entry in the `JUBE\_parameters'' configuration. An example of such an entry is: \texttt{srun -N 1 -n 128 -c 1 --exact python}, where \texttt{srun} is the command to initiate a slurm step within an existing allocation, \texttt{-N} specifies the number of nodes, \texttt{-n} defines the number of MPI processes, \texttt{-c} denotes the number of cores per process assigned to this step, and \texttt{--exact} tells the scheduler to assign only the previously specified resources to this slurm step. This configuration is detailed within the L2L execution script. %L2L has the capability to efficiently utilize a vast amount of resources distributed among the inner loop individuals.

%An alternative method involves requesting an independent allocation of resources for each individual. This option is advantageous when the resources required within the inner loop are substantial or when execution times are expected to vary across different parameters, thereby aiming to minimize idle resources.

%In this work, we opted for the first approach, as L2L demonstrated the capability to efficiently utilize a vast amount of resources distributed among the inner loop individuals.

\subsection{Fitting the transfer function}
\label{subsec:TransfFuncFit}

\begin{figure}[ht!]
\centering
\begin{subfigure}{.5\textwidth}
  \centering
  \includegraphics[width=1\linewidth]{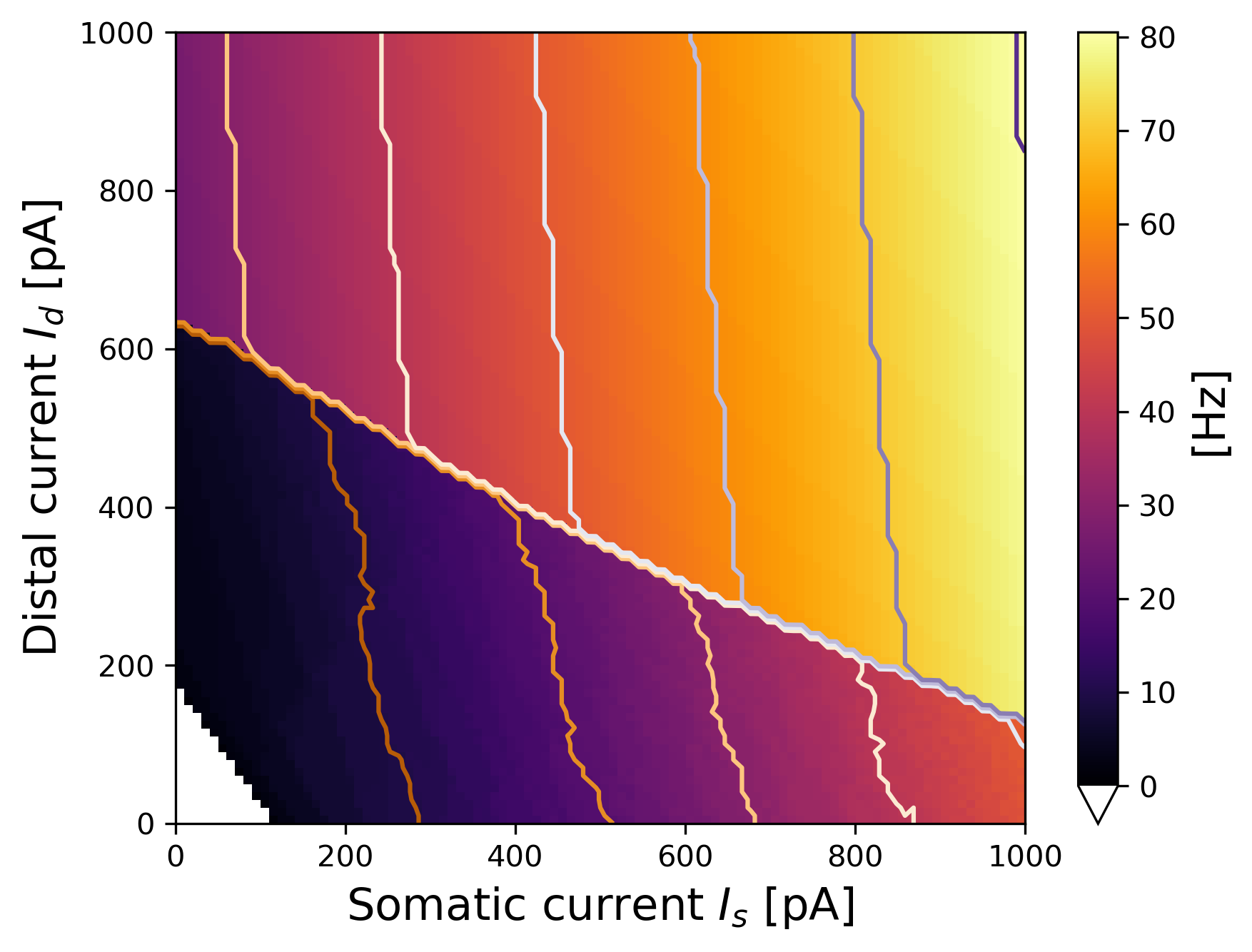}
  \caption{}
\label{fig:Simulated_Firing_Rates_cropped}
\end{subfigure}%
\begin{subfigure}{.5\textwidth}
  \centering
  \includegraphics[width=.82\linewidth]{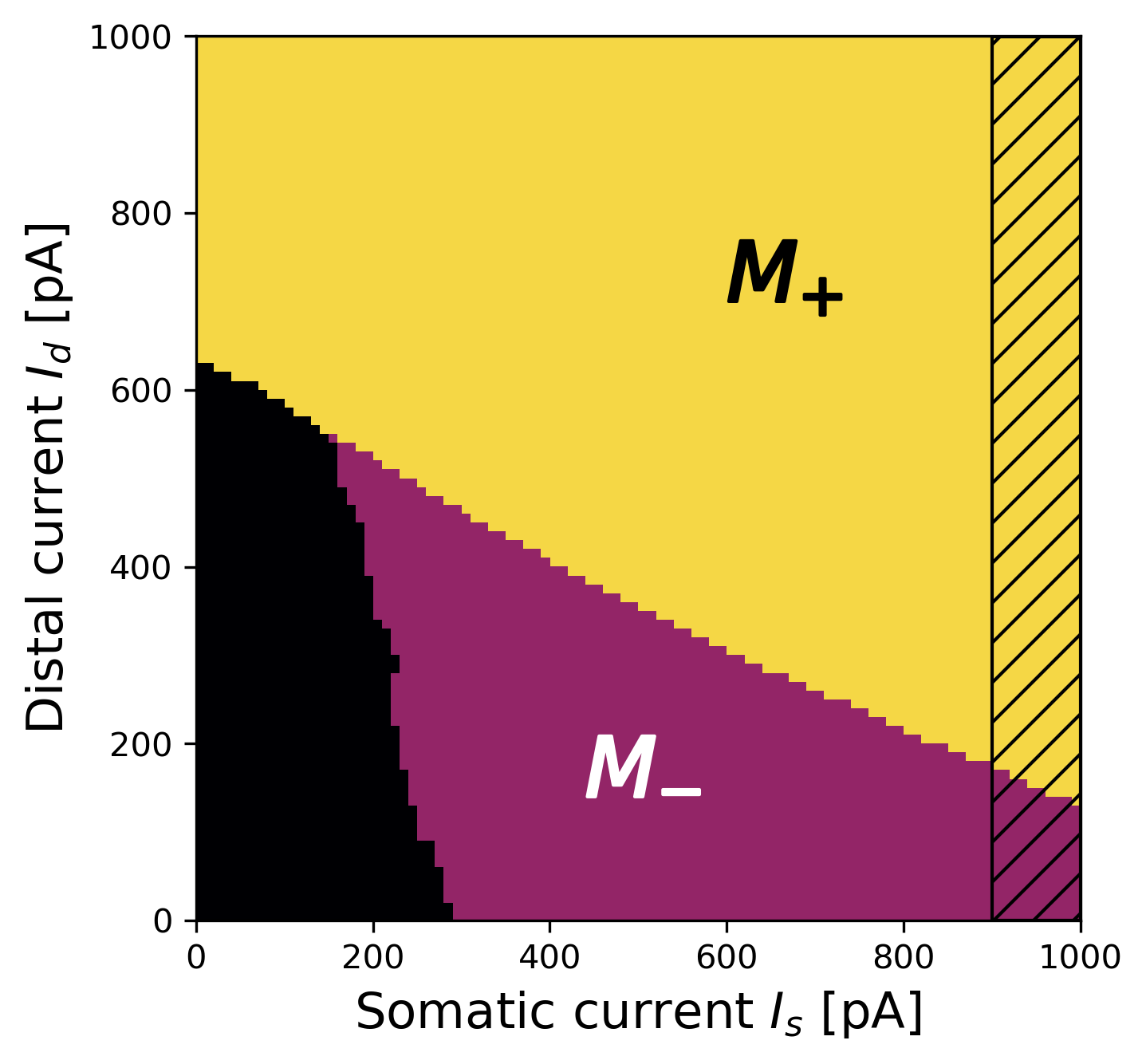}
  \caption{}
  \label{fig:mask_m_10Hz}
\end{subfigure}
\caption{Search for approximating planes. a) Representation of $\nu(I_s,I_d)$, the firing rate of the two-compartment Ca-AdEx spiking neuron in response to combinations of somatic ($I_s$) and distal ($I_d$) currents. b) Algorithmic identification of $M_{+}$ and $M_{-}$ regions from spiking simulation results.
}
\label{fig:thetaPlanesMethods_1}
\end{figure}

Figure \ref{fig:thetaPlanesMethods_1}.a illustrates $\nu(I_s,I_d)$, the firing rate of the exemplary two-compartment Ca-AdEx neuron identified by the evolutionary search algorithm (see Section \ref{sec:ExemplaryParams} for its parameters) in response to various combinations of constant somatic and distal currents. The regularity observed in the contour lines of equal firing rate suggests the potential for simplified approximate representations of the transfer function. This section outlines the method employed to derive such an approximation. Two distinct regions of low and high firing rate are discernible in Figure \ref{fig:thetaPlanesMethods_1}.a, seemingly demarcated by a straight line. Hereafter, we use the index $i \in \{-, +\}$ to denote the regions of lower or higher firing rates, respectively. In the $+$ region, contour levels of equal firing rate appear to be linear, parallel, and evenly spaced, indicating that the transfer function could be approximated by a plane. For each ($I_s$,$I_d$) pair, the simulation identifies the activation of the High Voltage dependent \Ca channel, resulting in a Boolean mask $M_{+}(I_s,I_d)$ that delineates the activation region associated with high firing rates (refer to Figure \ref{fig:thetaPlanesMethods_1}.b).

Fitting planes $\nu_{+}$ are defined by
\begin{equation}
\nu_{+}(I_s,I_d) =  a_{+} I_s + b_{+} I_d + d_{+}
\end{equation} 
and their parameters $(a_+,b_+,d_+)$ have been  identified in this work using the  \textit{LinearRegression} class from the \textit{sklearn.linear\_model} Python module (release 1.0.2). The same procedure returns the plane fitting the region of low activity $M_{-}$  (i.e., the lower part of Figure \ref{fig:thetaPlanesMethods_1}.a), where the contour lines are also approximately linear and evenly spaced for firing rates above a threshold $\nu_{low}$. 

The selection of an appropriate $\nu_{low}$ frequency is motivated by the need to model the learning advantages associated with apical amplification mechanisms, particularly in scenarios where external stimuli change at a fast rate. For instance, in a typical real-world scenario, sustaining a video rate of more than $20 frames/s$ is necessary, corresponding to an exposure to a stable perception lasting less than $50ms$. $\tau_{STDP}=20ms$ is a commonly chosen duration for an STDP mechanism that captures either correlations (multiplicative STDP) or causal influence (additive STDP) between a presynaptic neuron (\textit{pre}) and a postsynaptic neuron (\textit{post}), with the pair of nearest spikes occurring at $t_{post}$ and $t_{pre}$. Capturing even a single STDP-induced synaptic modification requires a minimum firing rate of $\nu_{low}>10 Hz$ and an exposure duration greater than $\simeq 50 ms$. In this context, a single synaptic modification event would typically be induced with $t_{post}-t_{pre}>2.5 \cdot \tau_{STDP}$. Therefore, for neurons capable of reaching significantly higher firing rates, capturing the regime of lower firing rates with extreme precision is not critical.

The $M_{-}(I_s,I_d)$ Boolean mask is defined by the points where the simulation indicates that $M_{+}(I_s,I_d) == \textit{false AND } \nu(I_s, I_d)>\nu_{low}$. 
The search for fitting planes can be done in the $M_{-}$ region, employing the same algorithm used for $M_+$, producing a $\nu_{-}(I_s,I_d)$ approximating plane.

\begin{figure}[ht!]
\centering
\begin{subfigure}{.45\textwidth}
  \centering
\includegraphics[width=1.\linewidth]{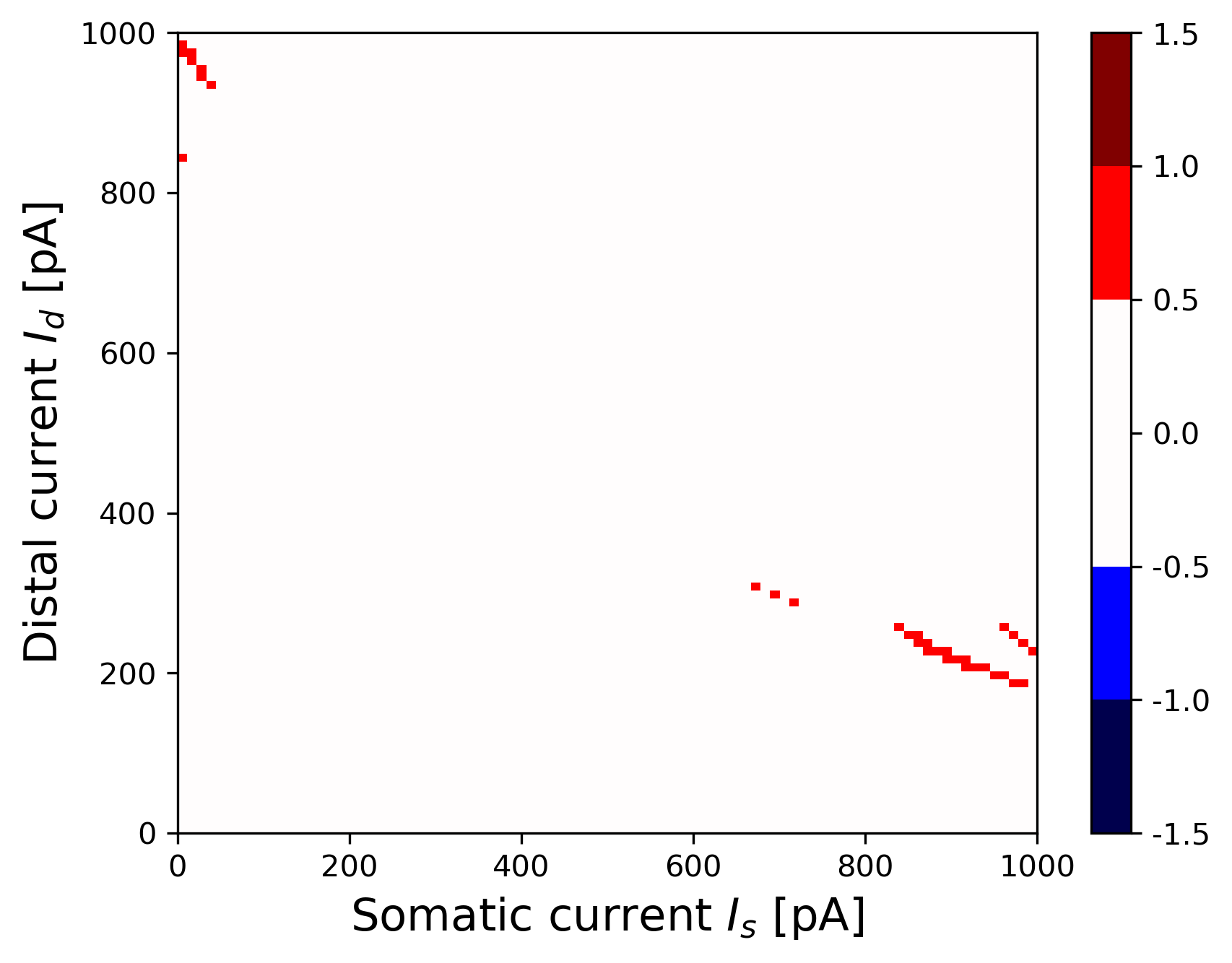}
  \caption{}
  \label{fig:diff2}
\end{subfigure}
\begin{subfigure}{.45\textwidth}
  \centering
\includegraphics[width=1.\linewidth]{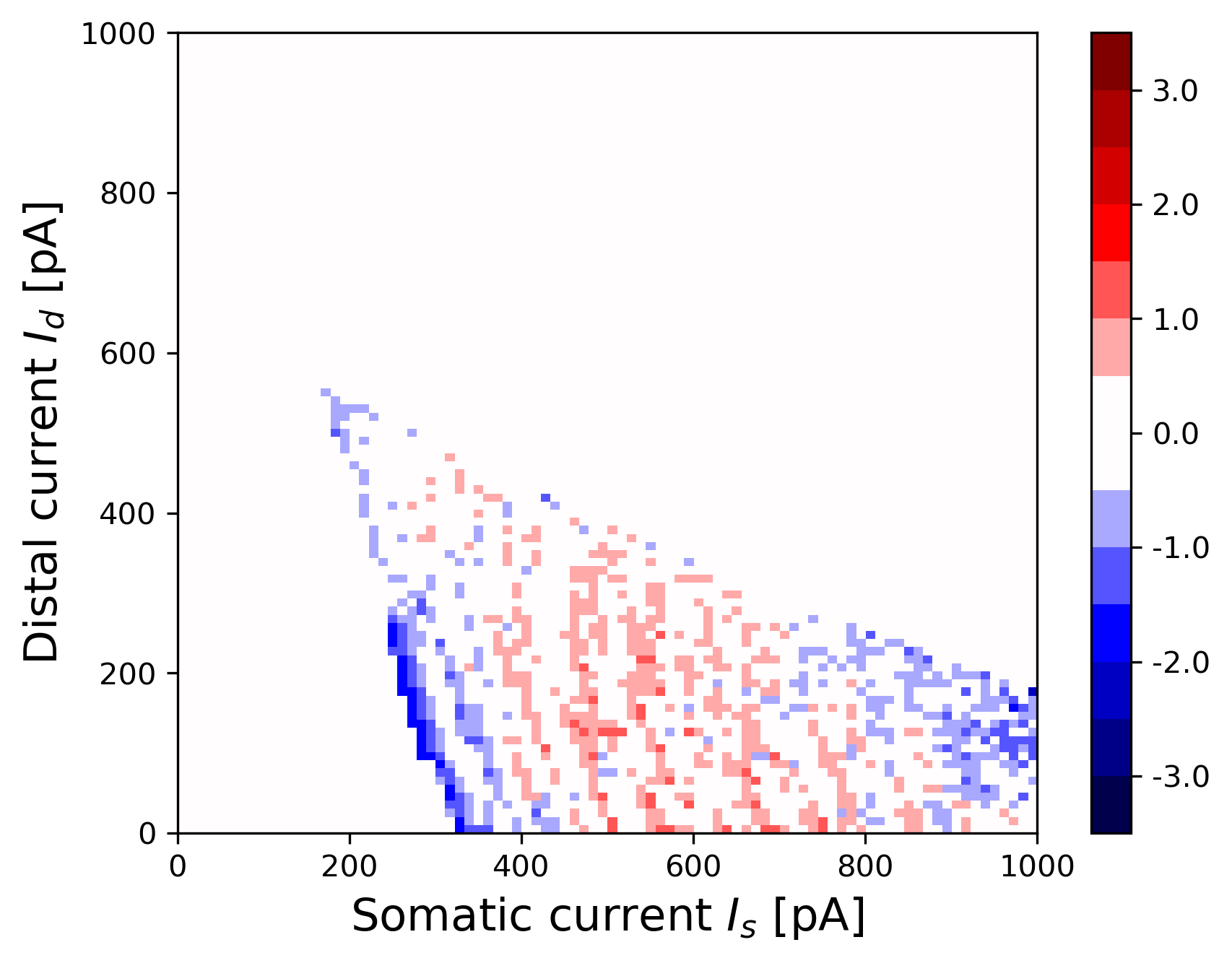}
  \caption{}
  \label{fig:fit1_10}
\end{subfigure}%
\caption{Errors of fitting planes (Hz). Panel a) $M_{+}$ region: $\nu_{+}-\nu$.  b) $M_{-}$ region: $\nu_{-}-\nu$.}
\label{fig:thetaPlanesError}
\end{figure}

To mitigate potential non-linearity at the boundaries of the region of interest, the Boolean masks excludes the region $I_s>I_{th}$, (illustrated as a dashed band in Figure \ref{fig:thetaPlanesMethods_1}.b).
Figure \ref{fig:thetaPlanesError}.a and \ref{fig:thetaPlanesError}.b show the error (in Hz) between the planar fits $\nu_{-,+}(I_s,I_d)$ and the simulated transfer functions $\nu(I_s,I_d)$. The discrepancy between the firing rates obtained from simulation of the two-compartment Ca-AdEx neuron and those predicted by the fitting plane is discretised into $0.5Hz$ intervals, reflecting that firing rates are measured over simulation periods lasting two seconds. 

The construction of $\nu_F(I_s,I_d)$, the simplified  description over the entire range of $I_s,I_d$ currents of the $\nu$ produced by spiking simulations, requires also a proper definition of the curve $I_d^H(I_s)$ that separates the $M_{+}$ from the  $M_{-}$ regions. $I_d^H(I_s)$ is the amount of distal current required to trigger a high firing regime ($H$)  given a fixed value of somatic $I_s$ current.

\begin{figure}[ht!]
\centering
\begin{subfigure}{.45\textwidth}
  \centering
  \includegraphics[width=1.\linewidth]{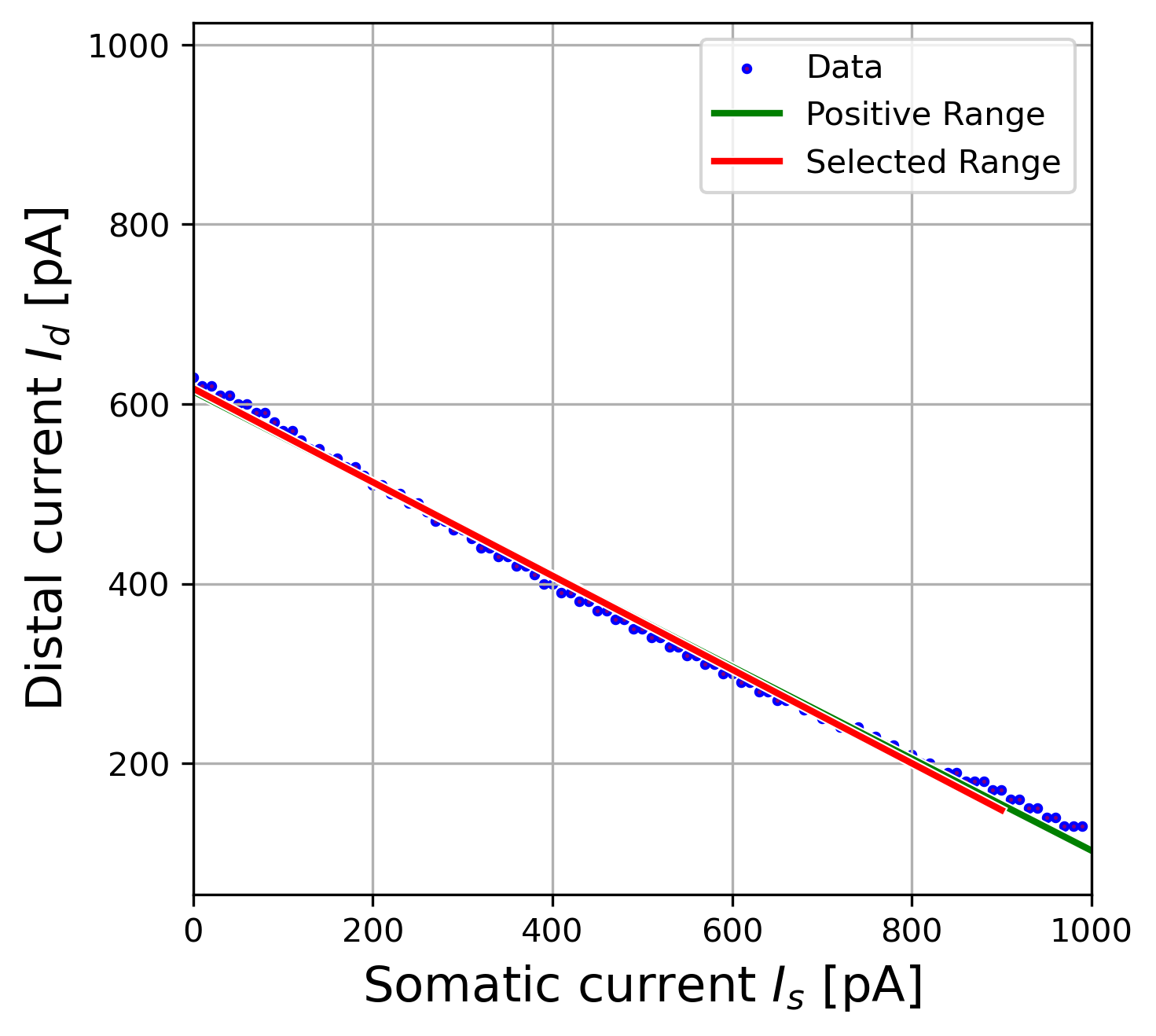}
  \caption{}
  \label{fig:sub21}
\end{subfigure}
\caption{Linearity of the separation between the high activity $M_{+}$ region and the lower activity $M_{-}$ region. (red line): $I_{d,F}^H(I_s)$ linear fit.}
\label{fig:thetaPlanesMethods_3}
\end{figure}

Figure \ref{fig:thetaPlanesMethods_3} illustrates that the linear fit  of the data representing the boundary between $M_+$ and $M_-$ leads to the definition of the parameters $\theta_m^H$ , the slope of the fitting line, and $\theta_q^H$, its offset.  The resulting approximating line is expressed as: 
\begin{equation}
    I_{d,F}^H(I_s) = \theta_{m}^{H}I_s + \theta_{q}^H
\end{equation}

Finally, the rheobase of the fitting function is defined by the combinations of currents that satisfy the condition $\nu_-(I_s, I_d) = 0$, this results in the line:
\begin{equation}
    I_{d,F}^\rho(I_s) = \theta_{m}^{\rho}I_s + \theta_{q}^\rho
\end{equation}

In summary, three planes ($\nu_0=0$,  $\nu_{-}(I_s,I_d)$ and $\nu_{+}(I_s,I_d)$) are identified by the algorithm to approximate the activity in each region. The active approximated domain is limited/bounded by:
\begin{equation}
    \Theta_H(I_s, I_d) = \Theta( I_d - I_{d,F}^{H}(I_s))
\end{equation}
The passive approximated domain is given by the product of two $\Theta$s, namely:
\begin{equation}
    \Theta_{\rho}(I_s, I_d) = \Theta(I_d - I_{d,F}^{\rho}(I_s))
\end{equation}
\begin{center}
and
\end{center}
\begin{equation}
    \Theta(- I_d + I_{d,F}^{}(I_s)) = (1 - \Theta_H(I_s, I_d)) 
\end{equation}
Finally, the fitting function that spans the entire domain, as determined by the algorithm, is:
\begin{equation}\label{fr_plane}
    \nu_{F}(I_s, I_d; \nu)  = \Theta_{\rho}(1- \Theta_H) \cdot \nu_{-} +  \Theta_H \cdot \nu_{+}
\end{equation}

This is referred to as \textit{ThetaPlanes} in the following.

\begin{table}[H]
\caption{Parameters of the \textit{ThetaPlanes} piece-wise linear approximating function. See Section \ref{sec:ExemplaryParams} for representative values.}
\label{tab:ThetaPlanesParameters}
\begin{center}
\begin{tabular}[t]{|c|c|}
\hline
\multicolumn{2}{|c|}{\textbf{$\nu_{+}$ plane, apical amplification region}}\\ 
\hline
$a_{+}$  & {Hz/pA}\\
\hline
$b_{+}$  & {Hz/pA}\\
\hline
$d_{+}$  & {Hz/pA}\\
\hline
\multicolumn{2}{|c|}{\textbf{$\nu_{-}$ plane, lower firing rate region}}\\ 
\hline
$a_{-}$  & {Hz/pA}\\
\hline
$b_{-}$  & {Hz/pA}\\
\hline
$d_{-}$  & {Hz/pA}\\
\hline
\multicolumn{2}{|c|}{\textbf{$I_{d,F}^H(I_s)$, line of separation between regions}}\\ 
\hline
$\theta_{m}^H$  & -\\
\hline
$\theta_{q}^H$  & pA\\
\hline
\multicolumn{2}{|c|}{\textbf{$I_{d,F}^\rho(I_s)$, rheobase line}}\\ 
\hline
$\theta_{m}^{\rho}$  & -\\
\hline
$\theta_{q}^{\rho}$  & pA\\
\hline
\end{tabular}
\end{center}
\end{table}

%changing parameters to support brain-state specific apical-amplificatio, apical-isolation and apical-drive (PSP,EP)
\subsection{Modulating the apical-amplification, -isolation and -drive regimes}
\label{subsec:DifferentApicalRegimesMethods}
\begin{figure}[ht!]
\centering
\begin{subfigure}{1.\textwidth}
  \centering
  \includegraphics[width=0.5\linewidth]{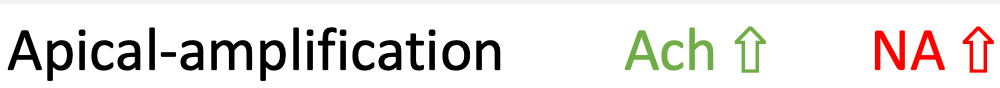}
\end{subfigure}
\begin{subfigure}{.3\textwidth}
  \centering
  \includegraphics[width=0.9\linewidth]{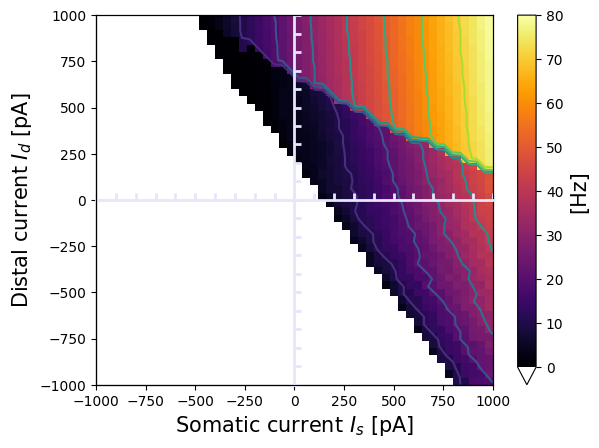}
  \caption{}
\end{subfigure}%
\begin{subfigure}{.3\textwidth}
  \centering
  \includegraphics[width=0.9\linewidth]{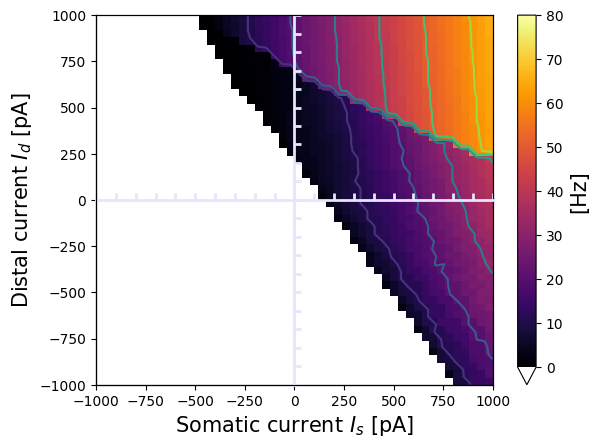}
  \caption{}
\end{subfigure}%
\begin{subfigure}{.3\textwidth}
  \centering
  \includegraphics[width=0.9\linewidth]{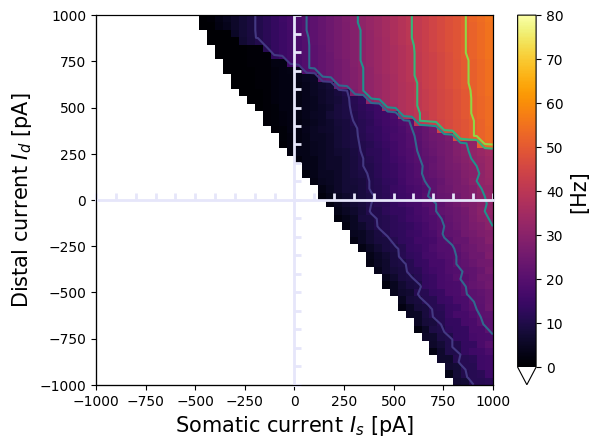}
  \caption{}
\end{subfigure}
\vspace{5pt}
\hrule height 1pt \relax
\vspace{5pt}
\begin{subfigure}{1.\textwidth}
  \centering
  \includegraphics[width=0.5\linewidth]{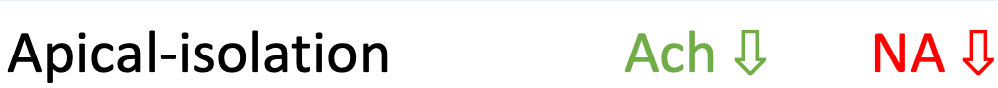}
\end{subfigure}
\begin{subfigure}{.3\textwidth}
  \centering
  \includegraphics[width=0.9\linewidth]{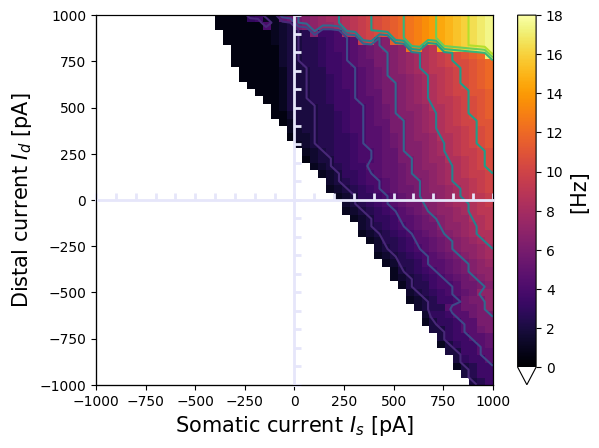}
  \caption{}
\end{subfigure}%
\begin{subfigure}{.3\textwidth}
  \centering
  \includegraphics[width=0.9\linewidth]{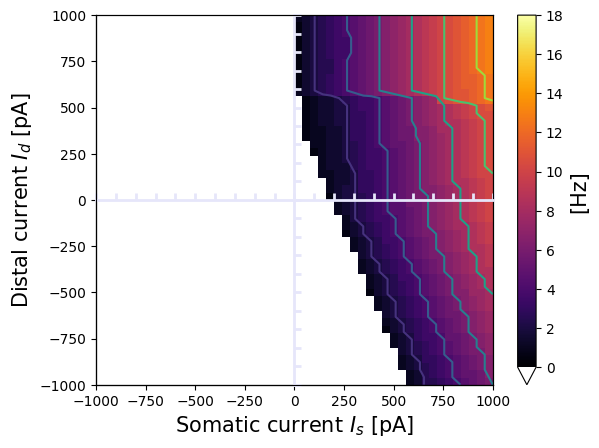}
  \caption{}
\end{subfigure}%
\begin{subfigure}{.3\textwidth}
  \centering
  \includegraphics[width=0.9\linewidth]{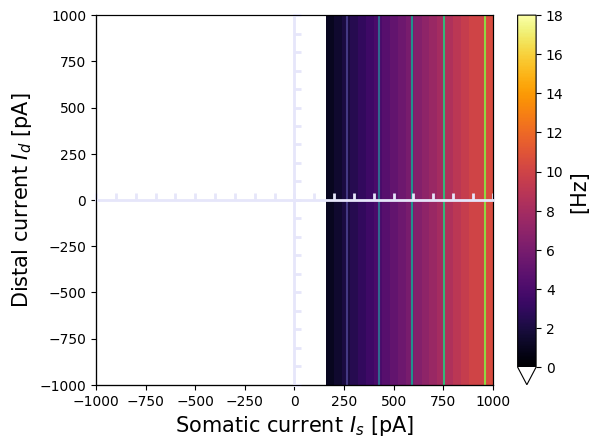}
  \caption{}
\end{subfigure}
\vspace{5pt}
\hrule height 1pt \relax
\vspace{5pt}
\begin{subfigure}{1.\textwidth}
  \centering
  \includegraphics[width=0.5\linewidth]{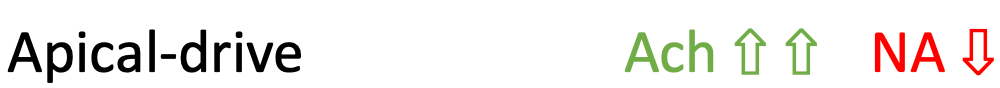}
\end{subfigure}
\begin{subfigure}{.3\textwidth}
  \centering
  \includegraphics[width=0.9\linewidth]{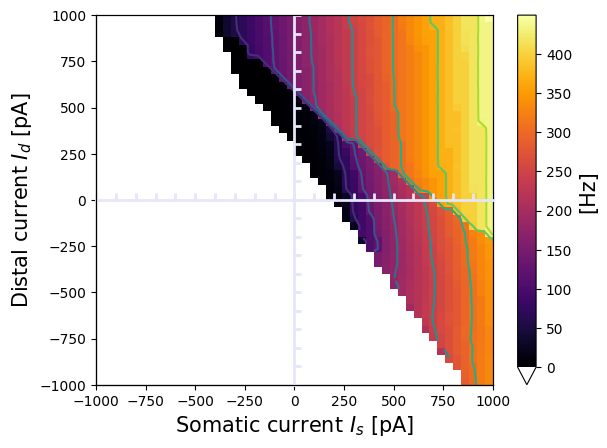}
  \caption{}
\end{subfigure}%
\begin{subfigure}{.3\textwidth}
  \centering
  \includegraphics[width=0.9\linewidth]{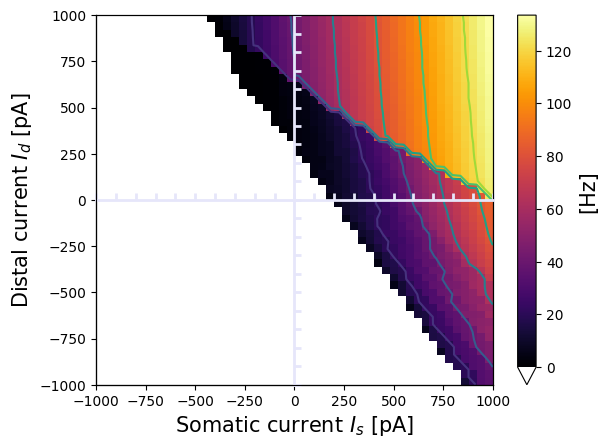}
  \caption{}
\end{subfigure}%
\begin{subfigure}{.3\textwidth}
  \centering
  \includegraphics[width=0.9\linewidth]{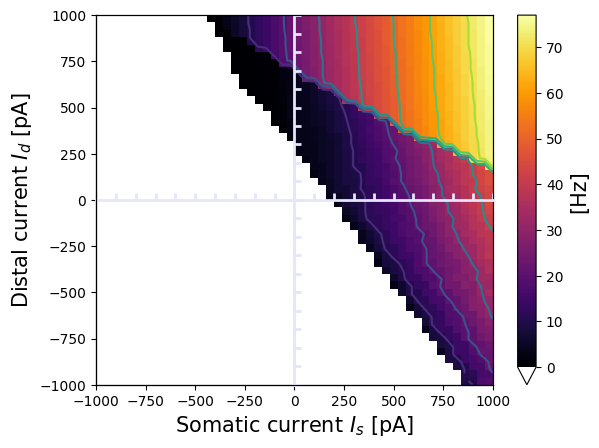}
  \caption{}
\end{subfigure}
\caption{Proxies for ACh and NA modulation. Inducing a range of apical-amplification -isolation and -drive like configurations from the same starting neuron. Parameters in table \ref{tab:ApicalRegimes-9-configurations}}
\label{fig:Apical-Amplification-Isolation-Drive-9Result}
\end{figure}

A few parameters serve as simulation proxies for the effects of neuromodulation, facilitating transitions to apical-isolation-like and apical-drive-like regimes or modulating the apical-amplification behavior. \cite{aruSiclariStorm2020} offers conceptual guidelines that have inspired the approach described here. As a proxy for ACh modulation, we consider the Spike Frequency Adaptation coefficient $b$ in eq.\ref{eq:AdEx}. Changes in excitability, associated with the level of NA, can be induced by, for example, altering the leakage reversal potential of the compartments (\cite{DandolaRebolloCasali2017}). Additionally, exploring the effect of a change in the conductance that connects the two compartments is of interest.
Figure \ref{fig:Apical-Amplification-Isolation-Drive-9Result} reports the results of the exploration in terms of neuromodulation for different brain states, using the parameters described in Table \ref{tab:ApicalRegimes-9-configurations}.

% Elena:
% Start commented code reproducing the figure of 9 candidates to apical-...
%to be used with separate panels

%\begin{comment}
%\hrule height h depth d width w \relax
%\end{comment}

\begin{table}[H]
\caption{Parameters to modulate the apical-amplification, -isolation, and -drive regimes. The apical-amplification configuration named \textit{a} is the one identified by the evolutionary search (see Section \ref{sec:ExemplaryParams} for its complete genome.) }
\label{tab:ApicalRegimes-9-configurations}
\begin{center}
\begin{tabular}[t]{|c|c|c|c|c|c|c|c|c|c|c|c|c|c|c|}
\hline
\multicolumn{5}{|c|}{Apical-amplification} &
\multicolumn{5}{c|}{Apical-isolation} &
\multicolumn{5}{c|}{Apical-drive} \\
\hline
Panel & $b$ & $g_C$ & $E_L^d$ & $E_L^s$ & Panel & $b$ & $g_C$ & $E_L^d$ & $E_L^s$ & Panel & $b$ & $g_C$ & $E_L^d$ & $E_L^s$\\
\hline
$a$ & 40 & 1 & -53 & -63 & $d$ & 200 & 1 & -58 & -68 & $g$ & 0 & 1 & -53 & -68 \\
$b$ & 50 & 1 & -53 & -63 & $e$ & 200 & 0.3 & -58 & -68 & $h$ & 20 & 1 & -53 & -68 \\
$c$ & 60 & 1 & -53 & -63 & $f$ & 200 & 0 & -58 & -68 & $i$ & 40 & 1 & -53 & -68\\

\hline
\end{tabular}
\end{center}
\end{table}

\subsection{Support for multi-compartment neurons in NEST}
\label{subsec:SupportInNEST}

To leverage existing technology for the efficient simulation of recurrently connected spiking neural networks, we have integrated a general multi-compartment (MC) modeling framework into NEST. Generally, MC models can be represented as
\begin{equation}\label{eq:MC}
    C^i \frac{dV^i}{dt} = g_L^i (E_L^i - V^i) + \sum_{c \in \mathcal{C}^i} I_c^i(\mathbf{y}_c^i, V^i) + \sum_{r \in \mathcal{R}^i} I_r^i(\mathbf{y}_r^i, V^i, S_r^i) + \sum_{j \in \mathcal{N}^i} g_C^{ij} (V^j - V^i),
\end{equation}
where $V^i$ denotes the membrane potential in compartment $i$, $C^i$ its capacitance, $g_L^i$ its leak conductance and $E_L^i$ the leak reversal potential. An arbitrary set $\mathcal{C}^i$ of ion channels may be present in compartment $i$. Their current $I_c^i(\mathbf{y}_c^i, V^i)$ depends on the local membrane potential and a set of channel state variables $\mathbf{y}_c^i$. Similarly, an arbitrary set $\mathcal{R}^i$ of synaptic receptors can exists, whose current may depend on state variables $\mathbf{y}_r^i$, the membrane potential, and the presynaptic input spike train $S_r^i$. Finally, the compartment $i$ is coupled to its neighbours $\mathcal{N}^i$ through a coupling conductance $g_C^{ij}$. Due to the conservation of current, the coupling is symmetric, i.e. $g_C^{ij} = g_C^{ji}$. By identifying the compartments with the nodes of a graph and the neighbour couplings with the edges, the MC model is always a tree graph.

In simulation tools for detailed biophysical models, the continuous cable model of neuronal morphology is discretized spatially through the second-order finite difference approximation \cite{Carnevale2006}, and the resulting system of equations takes the form of \eqref{eq:MC}. The number of compartments, or inversely their separation, is often chosen based on the electrotonic length constant. At a more abstract level, simplified multi-compartment (MC) models with two or three compartments are frequently utilized to represent elementary aspects of dendritic computation, with the parameters of \eqref{eq:MC} being tuned by ad-hoc methods for the specific scientific problem under investigation \cite{Pinsky1994, Clopath2007, Naud2014}. Between these levels of detail, compartmental parameters can be derived from full morphologies through matrix algebra to simulate local computations \cite{Wybo2021}, or they can be explicitly tuned to replicate these computations \cite{Pagkalos2023}.

The compartmental model architecture in NEST accommodates all these use cases by offering API functionality that enables end users to directly set compartmental parameters and arrange them in a user-specified tree graph layout. Furthermore, it is designed to be straightforwardly extendable with ion channels and receptor currents at the C++ level. 

The system is discretised in time using the Crank-Nicolson scheme:
\begin{equation}
    C^i \frac{V^i(t+h) - V^i(t)}{h} = \frac{F^i(V^i(t)) + F^i(V^i(t+h))}{2},
\end{equation}
where $F^i$ represents right-hand side of \eqref{eq:MC}. It is important to note that this method is implicit in the voltage: $F^i(V^i(t+h))$ needs to be Taylor-expanded so that all terms containing $V^i(t+h)$ ($\forall i \in \text{MC}$) can be moved to the left-hand side. The resulting matrix equation is then solved efficiently through the Hines algorithm \cite{Hines1984}. For the state variables of ion channels and receptor currents, we use the widely used leap-frog scheme: a state variable $y$ is computed at $t+\frac{h}{2}$, and thus has this value in both $F^i(V^i(t))$ and $F^i(V^i(t+h))$. Conversely, to compute the time evolution of a state variables from $t+\frac{h}{2}$ to $t+\frac{3h}{2}$, the voltage $V^i(t+h)$ is taken to be constant over this time-step. 

If the state variable follows the general Hodgkin-Huxley formalism, i.e.
\begin{equation}\label{eq:svar}
    \frac{dy}{dt} = \frac{y_{\infty}(V) - y}{\tau_y(V)},
\end{equation}
the value at time $t+\frac{3h}{2}$ follows from integrating this equation as an initial value problem starting from $y(t+\frac{h}{2})$, which has the analytical solution:
\begin{equation}
\begin{aligned}
    &y(t+\frac{3h}{2}) = P \, y(t+\frac{h}{2}) + \left( 1 - P \right) \, y_{\infty}(V(t+h)), \\ 
    &\text{with} \\
    &P = \exp\left(-\frac{h}{\tau_y(V(t+h))}\right).
\end{aligned}
\end{equation}
For state variables that do not depend on the voltage, as is often the case for those governing the synaptic conductance after spike arrival, efficiency is enhanced by precomputing the propagator $P$.

\subsection{Detailed morphological neuron model}
To demonstrate the potential of the MC modeling framework, we integrated Ca-AdEx into a neuron model that also includes dendritic compartments with NMDA-driven non-linearities, based after an L5PC morphology. This morphology was taken from Hay et al. \cite{Hay2011} and implemented in NEAT \cite{Wybo2021}. 
We focused on the most important somatic Na$^+$ and K$^+$ channels (NaTa and Kv3.1) and opted for a passive dendritic membrane. The physiological parameters recommended by Major et al. \cite{Major2008} were adopted to replicate the amplitudes of glutamate-uncaging evoked NMDA-spikes in L5PC dendrites and somata, combined with a spine correction as in Rhodes et al. \cite{Rhodes2006}. 
Concretely, this meant a specific capacitance of 0.8 $\mu$F/cm$^2$, which was increased by a factor 1.92 to account for spine surface in dendrites with a radius smaller than .6 $\mu$m. The axial resistance was set at 100 $\Omega$$\times$cm for smooth dendrites, and 120 $\Omega$$\times$cm for spiny dendrites, while the specific membrane condutance was 100 $\mu$S/cm$^2$, and the leak reversal was fixed at -75 mV. 

This model was then simplified into a description with six distal apical and 8 distal basal compartments that received AMPA+NMDA as well as GABA synapses, in addition to the soma and a Ca-hotzone compartment located where the main apical trunk splits into multiple branches.
For technical reasons, all bifurcation sites in between any of those compartments are added automatically by the simplification procedure (Figure~\ref{fig:nmda}B, \cite{Wybo2021}).
The parameters of the reduced model that also featured in the Ca-AdEx optimization procedure (such as the leak and capacitance of soma and Ca-HZ compartments, as well as their coupling) were overwritten by those obtained through the optimization, and the other optimized parameters of the Ca- and AdEx-mechanisms where added as well.

The resulting model was then stimulated with input current steps (Figure~\ref{fig:nmda}C), the BAC-firing protocol (Figure~\ref{fig:nmda}D) and Poisson distributed synaptic inputs (Figure~\ref{fig:nmda}E,~F).
For the BAC-firing protocol, we used a somatic current step amplitude of 750 pA and a double exponential input current at the Ca-HZ compartment with $\tau_r = 1$ ms and $\tau_d = 10$ ms, and which further had a maximal amplitude of 1500 pA.
For the Poisson synaptic inputs, AMPA and GABA receptors were simulated as the product of a double exponential conductance window \cite{Rotter1999} $g$ and a driving force:
\begin{equation}\label{eq:ICOND}
i_{\text{syn}} = g \, (e_r-v), \hspace{4mm} \text{with} \hspace{4mm} g = w \, n(\tau_r, \tau_d) \, \left( e^{-t / \tau_d} - e^{-t / \tau_r} \right).
\end{equation}
Here, $e_r$ is the synaptic reversal potential, $\tau_r$ and $\tau_d$ are the synaptic rise and decay time constants, and $n$ a normalization constant that depends on $\tau_r$ and $\tau_d$ and normalizes conductance window $g$, so that its peak value is equal to the synaptic weight $w$.
AMPA rise and decay times were ${\tau_r = 0.2 \; \text{ms}}$, ${\tau_d = 3 \; \text{ms}}$ and AMPA reversal potential was ${e_r = 0 \; \text{mV}}$, whereas for GABA, we had ${\tau_r = 0.2 \; \text{ms}}$, ${\tau_d = 10 \; \text{ms}}$ and ${e_r = -80 \; \text{mV}}$. NMDA currents \cite{Jahr1990a} were implemented as:
\begin{equation}\label{eq:INMDA}
i_{\text{syn}} = g \, \sigma (v) \, (e_r-v)
\end{equation}
with ${\tau_r = 0.2 \; \text{ms}}$, ${\tau_d = 43 \; \text{ms}}$, and  ${e_r = 0 \; \text{mV}}$, while $\sigma (v)$ -- the channel's magnesium block -- had the form \cite{Behabadi2014}:
\begin{equation}
\sigma(v) = \frac{1}{1 + 0.3 \, e^{-0.1 \, v}}.
\end{equation}
The synaptic weight (i.e. maximum value of the conductance window) for the AMPA component of AMPA + NMDA synapses was set at 1 nS, and the maximal value of the NMDA window was twice that of the AMPA window (NMDA ratio of 2). GABA synapses also had a weight of 1 nS. While for the AMPA+NMDA synapses a multitude of Poisson input rates were probed as part of the scan (Figure~\ref{fig:nmda}F), the Poisson input rate to the GABA synapses was fixed at 20 Hz.

    % cm: float = 0.8 # Cm [uF/cm^2]
    % ra_smooth: float = 100. / 1e6 # Ra [MOhm*cm]
    % ra_spiny: float = 120. / 1e6  # Ra [MOhm*cm]

    % gl: float = 100. # dendritic leak conductance [uS/cm^2]
    % el: float = -75. # dendritic leak reversal [mV]

    % # spine correction from Rhodes et al. (2006)
    % sf: float = 1.92 # spine factor
    % r0: float = 0.6 # where spines occur [um] (radius < r0)

    % # channel densities simplified from Hay et al. (2011)
    % g_na: float = 1.71 * 1e6 # somatic sodium conductance [uS/cm^2]
    % e_na: float = 50. # sodium reversal [mV]

    % g_k: float = 0.766 * 1e6 # somatic potassium conductance [uS/cm^2]
    % e_k: float = -85. # potassium reversal [mV]

    % e_eq: float = -75. # somatic equilibirum potential [mV]
    % tau_s: float = 10. # somatic target membrane timescale [ms]

\section{Results}
\label{sec:Results}
\subsection{Dynamics of the two-compartment Ca-AdEx neuron}
\label{subsec:Ca-AdExEquations}
By combining the multi-compartment neuron equation (\ref{eq:MC}) with the AdEx equations (\ref{eq:AdEx}), considering the ion currents detailed in equations (\ref{eq:Ca_curr}) and (\ref{eq:K_curr}), the calcium concentration dynamics (\ref{eq:calcium_conc_Larkum}), and the BAP contribution, the dynamics of Ca-AdEx neuron (outside the refractory period) is described by:

\begin{equation}
    \label{eq:Ca-AdEx}
    \left\{
    \begin{array}{ll}
        \left\{
        \begin{array}{ll}
            C_m^s \frac{dV^s}{dt} &=
            -g_L^s(V^s-E_L^s)+g_L^s\Delta_T\exp\left(\frac{V^s-V_{th}^s}{\Delta_T}\right) +
            \\
            \\
            &\quad - g_e^s(t)(V^s-E_e^s)-g_i^s(t)(V^s-E_i^s) +
            \\
            \\
            &\quad - w + I_e^s - g_C(V^s-V^d) \\
            \\
        %\label{eq:somaCompAdaptation}
            \tau_w \frac{dw}{dt} &= a(V^s-E_L^s) +b\sum_{k}\delta (t-t_{k}) - w
            \\
            \\
        \end{array}
        \right.
        \\
        \\
        \begin{array}{ll}
        \quad C_m^d \frac{dV^d}{dt} &=
        -g_L^d(V^d-E_L^d) - g_e^d(t)(V^d-E_e^d)-g_i^d(t)(V^d-E_i^d) + 
        \\
        \\
        &\quad + I_{Ca} + I_{K_{Ca}} + w_{BAP}\sum_{k}\delta (t-(t_{k}+d_{BAP}))+
        \\
        \\
        &\quad + I_e^d + g_C(V^d-V^s)
        \end{array}
    \end{array}
    \right.
\end{equation}
A somatic spike event is triggered when $V^s >= V_{th}$, which defines the $t_k$ spike time. $V^s$ is set to the constant value $V_{reset}$ during  $t_k < t= <t_k+ t_{ref}$, while the distal compartment continues  to integrate the dynamics defined by equation (\ref{eq:Ca-AdEx}) during this period.

\begin{figure}[ht!]
  \centering
  \includegraphics[width=0.6\textwidth]{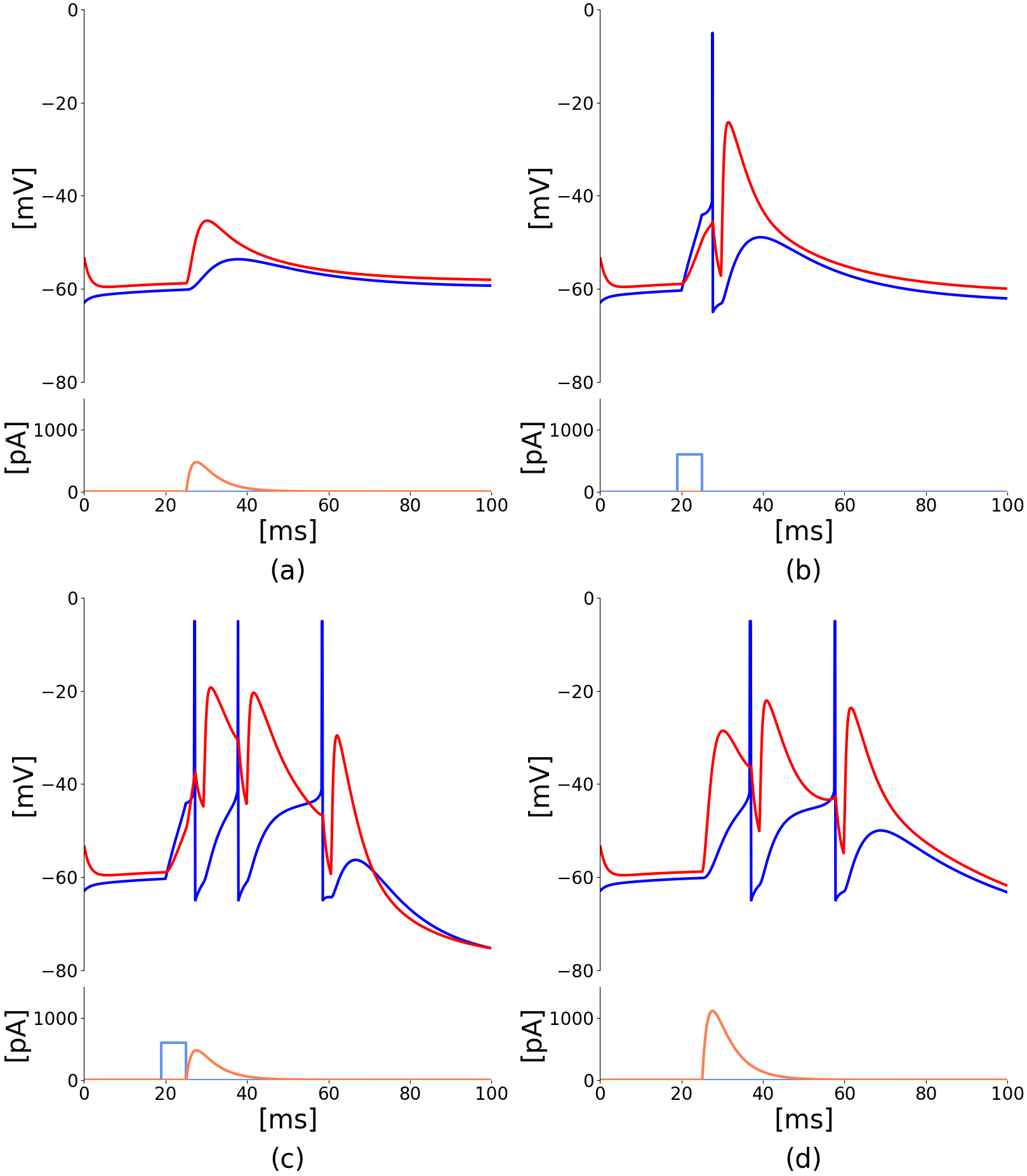}
  \caption{Response to pulse stimuli. 
  (a) a beta-shaped current injection of 950pA (peak amplitude) at the distal compartment produces a deflection of only 11mV at the soma without eliciting any spike;
  (b) a threshold current injection (550pA) at the soma evokes one single AP;
  (c) the combination of a threshold somatic current as in b) and an under-threshold distal current as in a), separated by an interval of 5ms, activates the BAC firing mechanism and evokes a burst of three APs;
  (d) to obtain a burst using only distal injection, a current of at least 1350pA is required. 
  All panels share scale bars and legend: in blue somatic membrane voltage; in red distal membrane voltage; in lightblue step somatic input; in orange beta-shaped distal input).
  }
  \label{fig:pulseProtocol}
\end{figure}

\subsection{Response to pulse stimuli}
Employing the L2L optimization framework to select neurons that best match our criteria led to the identification of the best fitting model. Figure \ref{fig:pulseProtocol} illustrates the behavior of the fitted model in response to input currents of short duration (a few milliseconds), according to the protocol for the \textit{pulse stimuli} task outlined in \ref{subsec:FitnessFunc}.
An under-threshold distal input, modeled as a beta function to mimic an excitatory postsynaptic potential (EPSP), slightly deflects the somatic membrane potential but does not trigger any spikes (\ref{fig:pulseProtocol}.a). A threshold somatic signal that evokes an action potential (AP) back-propagates through the axon, stimulating a \Ca influx that results in an increase in the distal membrane potential but is insufficient to initiate a calcium spike (\ref{fig:pulseProtocol}.b).

The concurrent application of the previous two input signals triggers a burst of two to three spikes at approximately $5-10$Hz: the back-propagating action potential, induced by the threshold somatic input, lowers the membrane potential in the Ca-HZ. When coupled with the under-threshold distal input, this facilitates the initiation of the calcium spike (\ref{fig:pulseProtocol}.c).
To generate a similar burst with solely distal input, a higher peak current value must be supplied, as demonstrated in the example of \ref{fig:pulseProtocol}.d.
When combining somatic and distal input currents, the distal current is introduced with a delay of $5$ms relative to the somatic one. Analyzing the neuron's performance concerning this delay is not covered in this work, but it will be considered for further optimization of the neuron.

%\subsection{Identification of interesting genomes}
%\label{subsec:SampleGenomes}
%Example of neural genome see Table \ref{tab:SampleGenomeTable}

%\begin{table}[H]
%\caption{Sample genome of a two-compartment neuron model optimized to express apical amplification}
%\label{tab:SampleGenomeTable}
%\begin{center}
%\begin{tabular}[t]{|c|c|}
%\hline
%\multicolumn{2}{|c|}{\textbf{Neuron Genome}}\\ 
%\hline
%$C_m$ & 200.0\\
%\hline
%$g_L$ & 10.0\\
%\hline
%$...$ & ...\\
%\hline
%\end{tabular}
%\end{center}
%\end{table}

%The use of the L2L optimization framework for the selection of neurons that best fits our criteria, leads to the identification of a few candidates. Among them, by visual inspection, we chose the one that best match our hypothetical target, described in section \ref{subsec:FitnessFunc}.

%\elena{Table of parameters for the selected neuron}.

%\elena{Match with target}

\begin{figure}[ht!]
\centering
\begin{subfigure}{.99\textwidth}
  \centering
  \includegraphics[width=1.\linewidth]{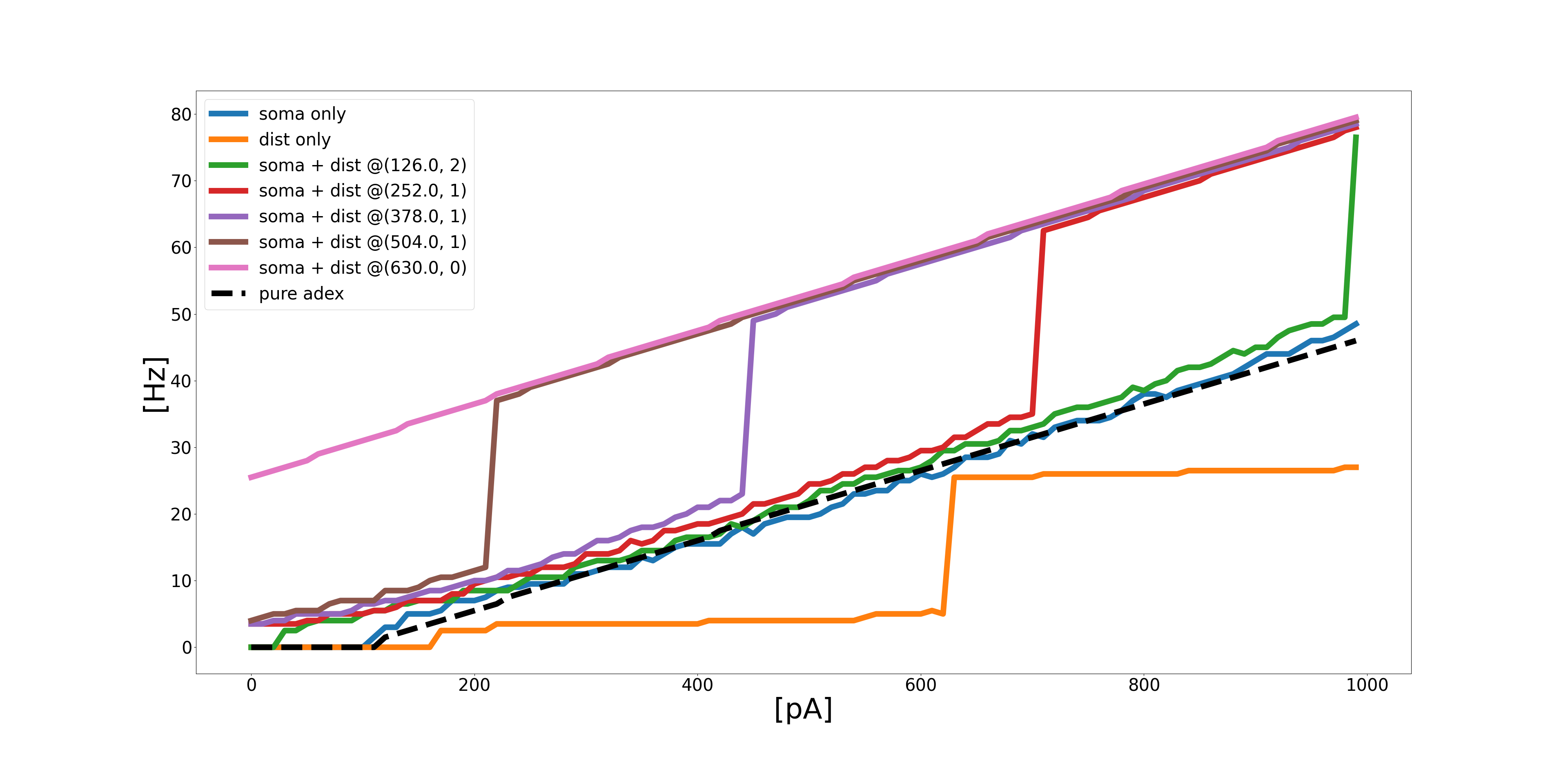}
  \caption{}
  \label{fig:sub1}
\end{subfigure}
%\begin{subfigure}{.3\textwidth}
%  \centering
%  \includegraphics[width=1.\linewidth]{Figures/ThetaPlanesFiringRates_points.png}
%  \caption{}
%  \label{fig:sub2}
%\end{subfigure}
\begin{subfigure}{.5\textwidth}
  \centering
  \includegraphics[width=1.\linewidth]{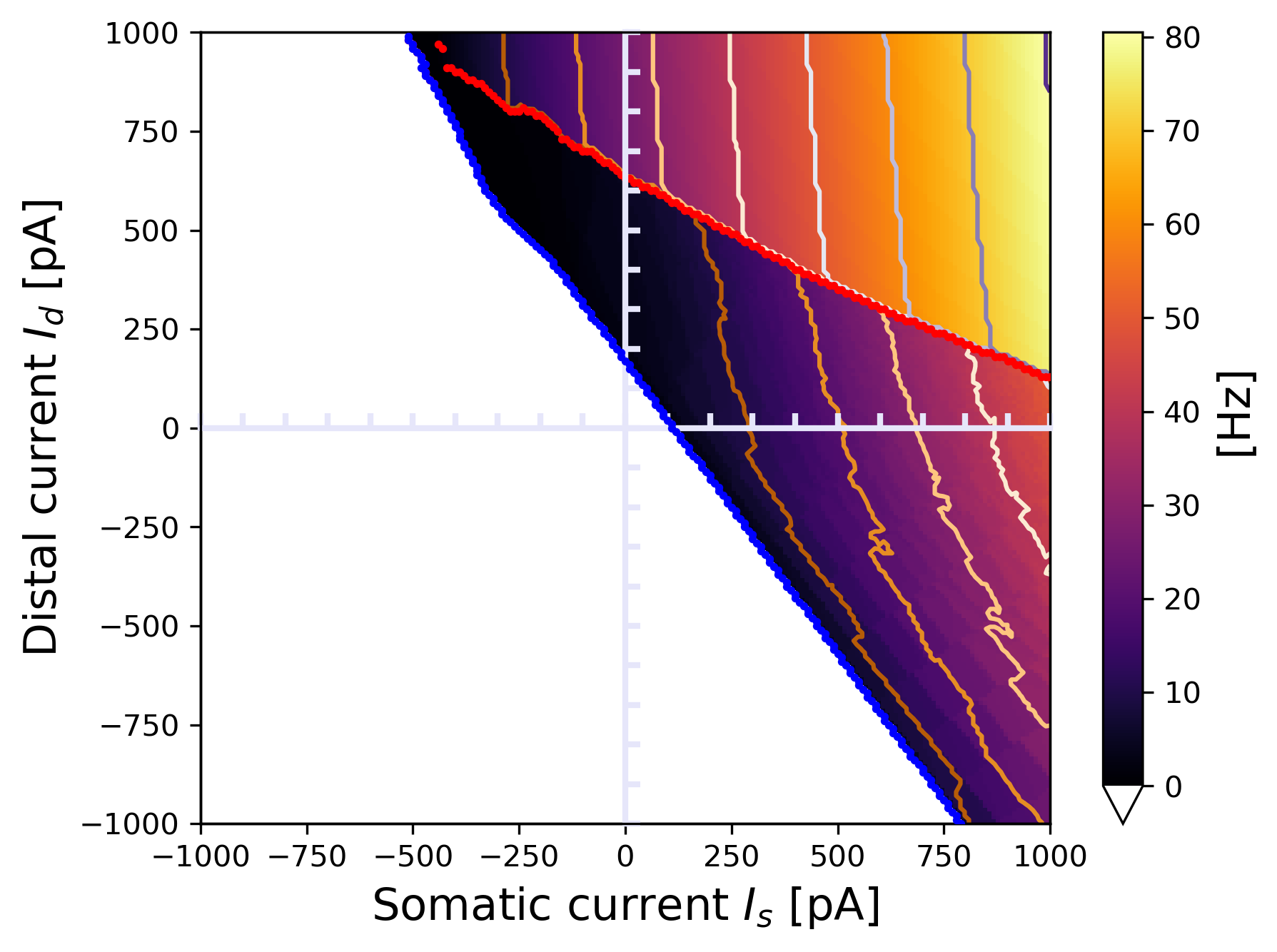}
  \caption{}
  \label{fig:sub3}
\end{subfigure}%
\begin{subfigure}{.5\textwidth}
  \centering
  \includegraphics[width=1.\linewidth]{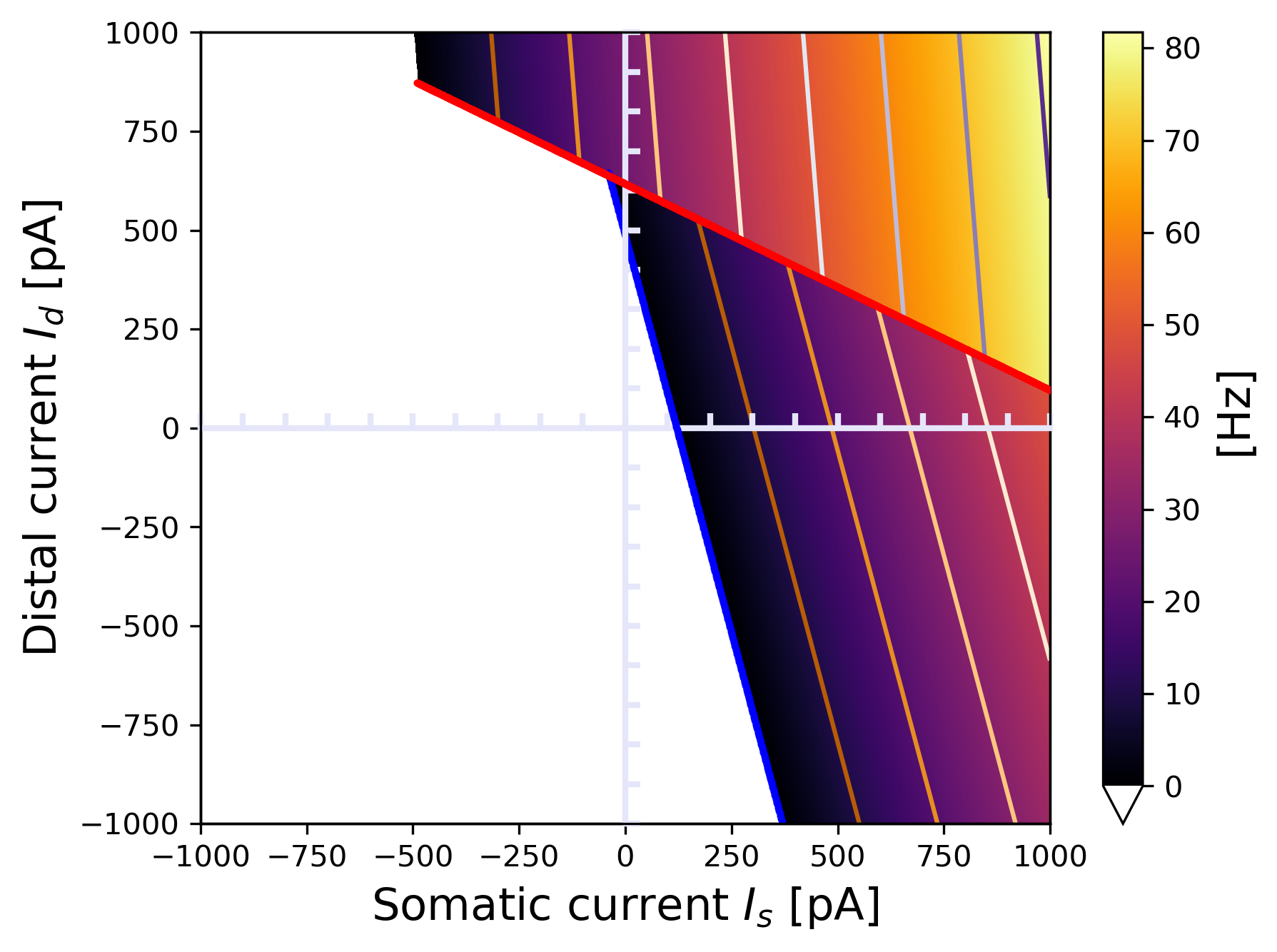}
  \caption{}
  \label{fig:sub4}
\end{subfigure}
\caption{Transfer function of selected neuron and it's approximation with ThetaPlanes. (a) current to rate response to DC inputs delivered to different compartments: pure somatic current (blue), pure distal current (orange), combination of somatic and distal current (other colored lines). The dashed black line represents the transfer function of the AdEx neuron used as the target reference for the fitness function.  
(b)Transfer function of the neuron in the 2-D plane defined by somatic and distal input DC currents; in blue the rheobase and in red the transition line between passive and active calcium regimes, respectively expressing a lower and an higher firing rate. 
(c) ThetaPlanes approximating the transfer function.}
\label{fig:bestNeu}
\end{figure}

\subsection{Response to prolonged stimulus: compact geometric description of the transfer function}
\label{subsec:TransfFuncDesc}

Figure \ref{fig:bestNeu} summarizes the primary characteristics of the selected neuron in response to the \textit{prolonged stimuli} task. Panel (a) illustrates the neuron's dynamics for specific distal and somatic input currents, as administered according to the guidelines detailed in section \ref{subsec:FitnessFunc}.
The orange line depicts the firing rate when the neuron is stimulated solely with a distal current. The activation of the BAC firing mechanism is indicated by the sharp increase in the firing rate observed at $630$pA in the orange line: beyond this threshold, even without somatic input, a dendritic calcium spike is initiated, and the neuron enters into the active regime, wherein the mechanism of apical amplification becomes apparent. The other curves illustrate the neuron's response when stimulated with a combination of currents injected into both the soma and the distal compartment. The visible jump in these curves corresponds to the neuron entering the active regime, a state reached when the combined effect of the two input currents is sufficient to trigger the calcium spike. As the value of the constant distal current increases, the transition to the active regime occurs at progressively lower somatic input currents. The blue line represents the scenario where $I_d=0$: neither the calcium spike nor the BAC firing mechanism is triggered, and within the analyzed range, the neuron behaves similarly to the pure AdEx model against which the two-compartment model has been fitted (indicated by the black dashed line).

Figure \ref{fig:bestNeu}.b displays the firing rate of the neuron when stimulated with combinations of somatic and distal currents ($\nu(I_s,I_d)$). Three distinct regions are identifiable: the area below the blue line, where the firing rate equals $0$ for every input current combination; an area of low firing rates situated between the blue and red lines; and an area of high firing rates above the red line, indicating the triggering of the calcium spike and the activation of the apical amplification mechanism (active regime). The blue line denotes the neuron's rheobase, while the red line signifies the transition from the passive to the active regime.
As discussed in the \nameref{sec:Methods} section, by examining the firing rate $\nu$ of the multi-compartment neuron in the plane of input somatic and distal currents $I_s$ and $I_d$, we can create a simplified model at a significantly higher level of abstraction. This is achieved through the definition of fitting planes: one for the apical amplification zone and another for the lower activity region of the neuron's transfer function.
%The Montecarlo search (see. Figure \ref{fig:bestNeu}.b), identifies triads of points that define fitting planes, one for the apical amplification zone and the other for the lower activity region of the neuron transfer function.

$\nu$ can be piece-wise by planes separated by lines, resulting in the \textit{ThetaPlanes} transfer function:

\begin{equation}\label{fitting_plane}
    ThetaPlanes(I_s, I_d; \nu)  = \Theta_{\rho}(1- \Theta_H) \cdot \nu_{-} +  \Theta_H \cdot \nu_{+}
\end{equation}

Figure \ref{fig:bestNeu}.c displays such approximating function.

Table \ref{tab:ThetaPlanesParameters} lists the parameters that define the \textit{ThetaPlanes} function: the $v_{-}(I_s,I_d)$ and $v_{+}(I_s,I_d)$ planes, the transition line to high firing rates, and the rheobase. The \textit{ThetaPlanes} configuration fitting the exemplary neuron is detailed in Section \ref{sec:ExemplaryParams}.

\subsection{Wakefulness, NREM and REM specific apical mechanisms}
\label{subsec:ResultsApicalRegimesMethods}

\begin{figure}[ht!]
\centering
\begin{subfigure}{.3\textwidth}
  \centering
  \includegraphics[width=0.9\linewidth]{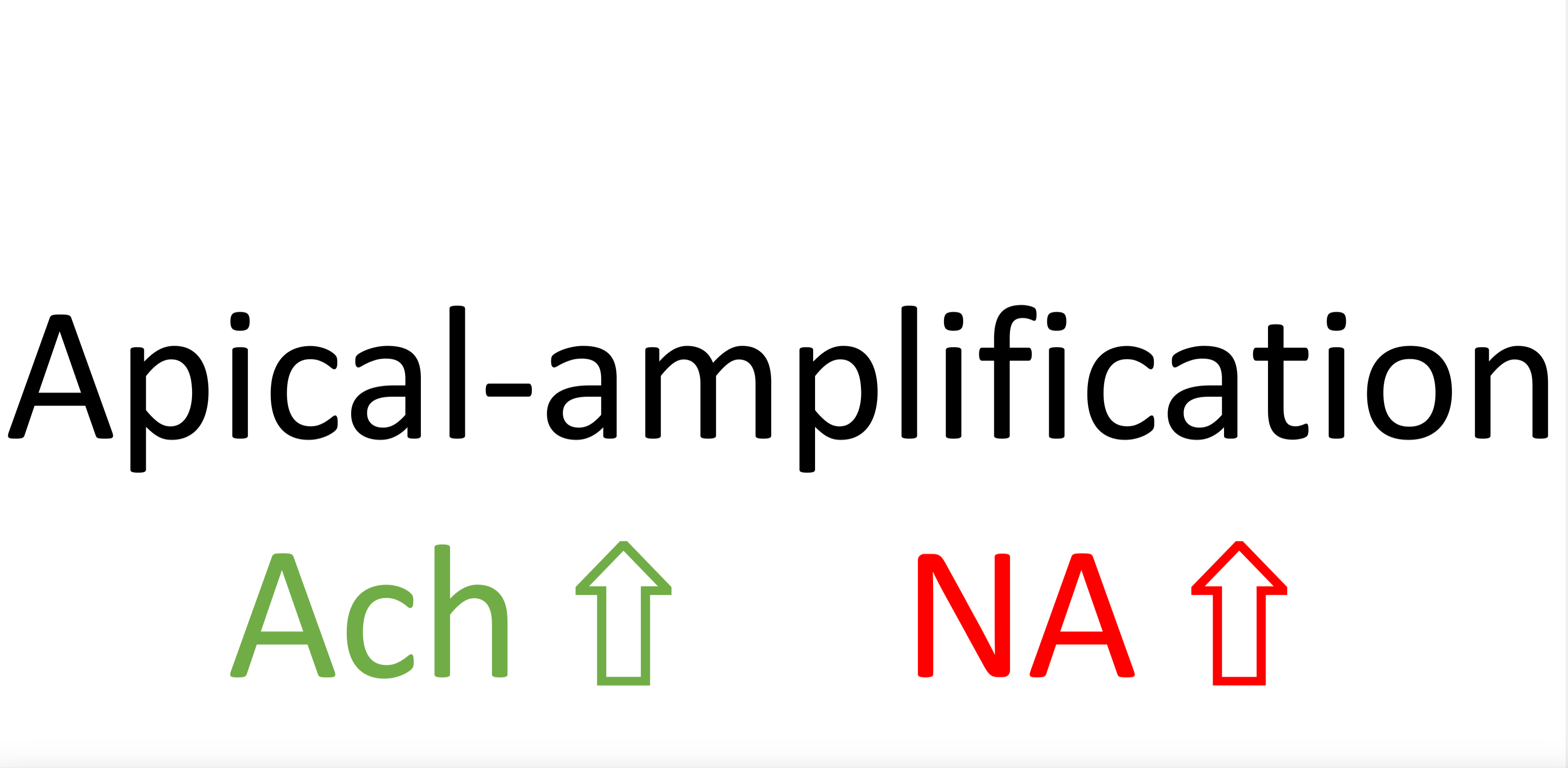}
\end{subfigure}%
\begin{subfigure}{.3\textwidth}
  \centering
  \includegraphics[width=0.9\linewidth]{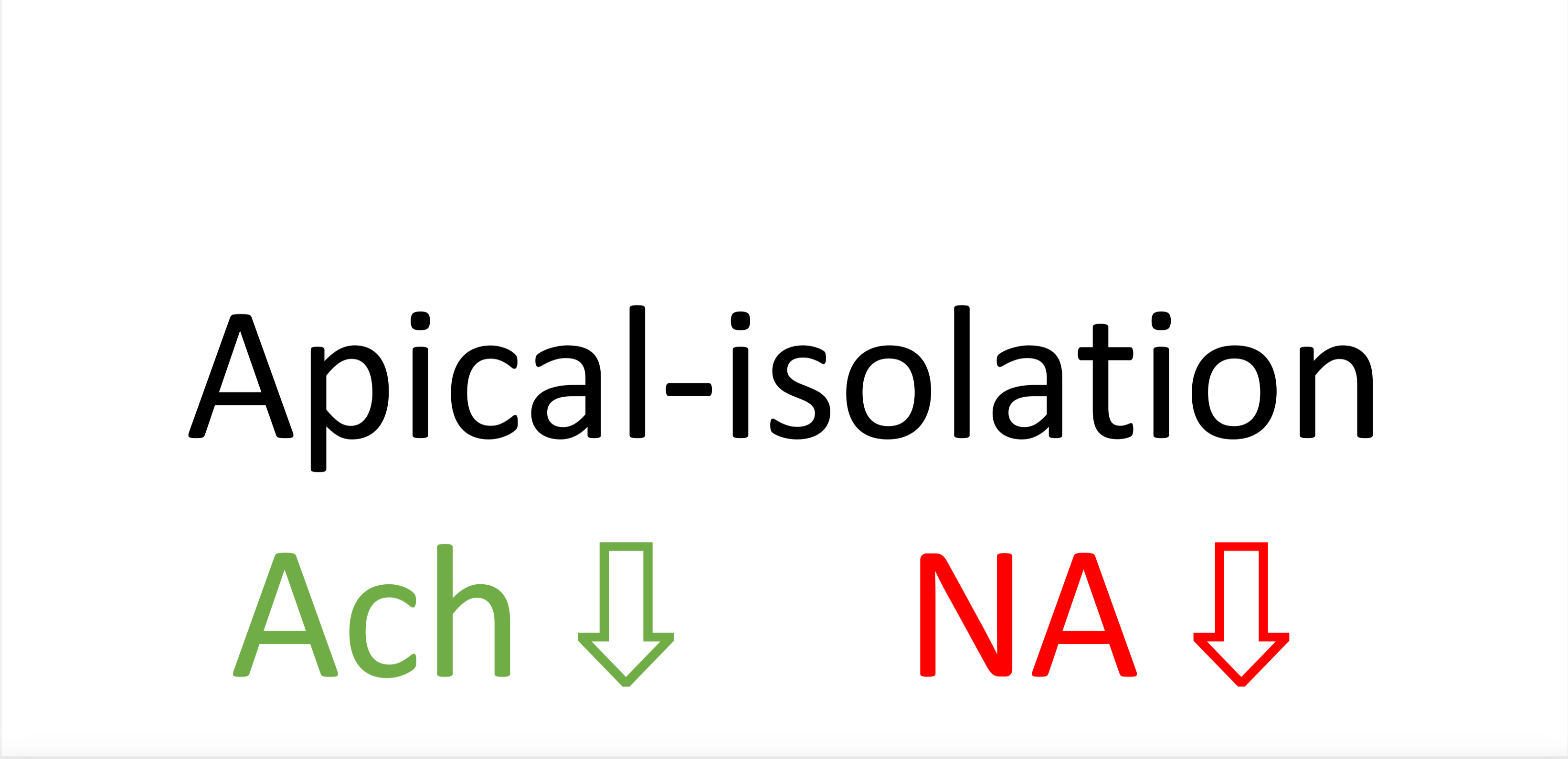}
\end{subfigure}%
\begin{subfigure}{.3\textwidth}
  \centering
  \includegraphics[width=0.9\linewidth]{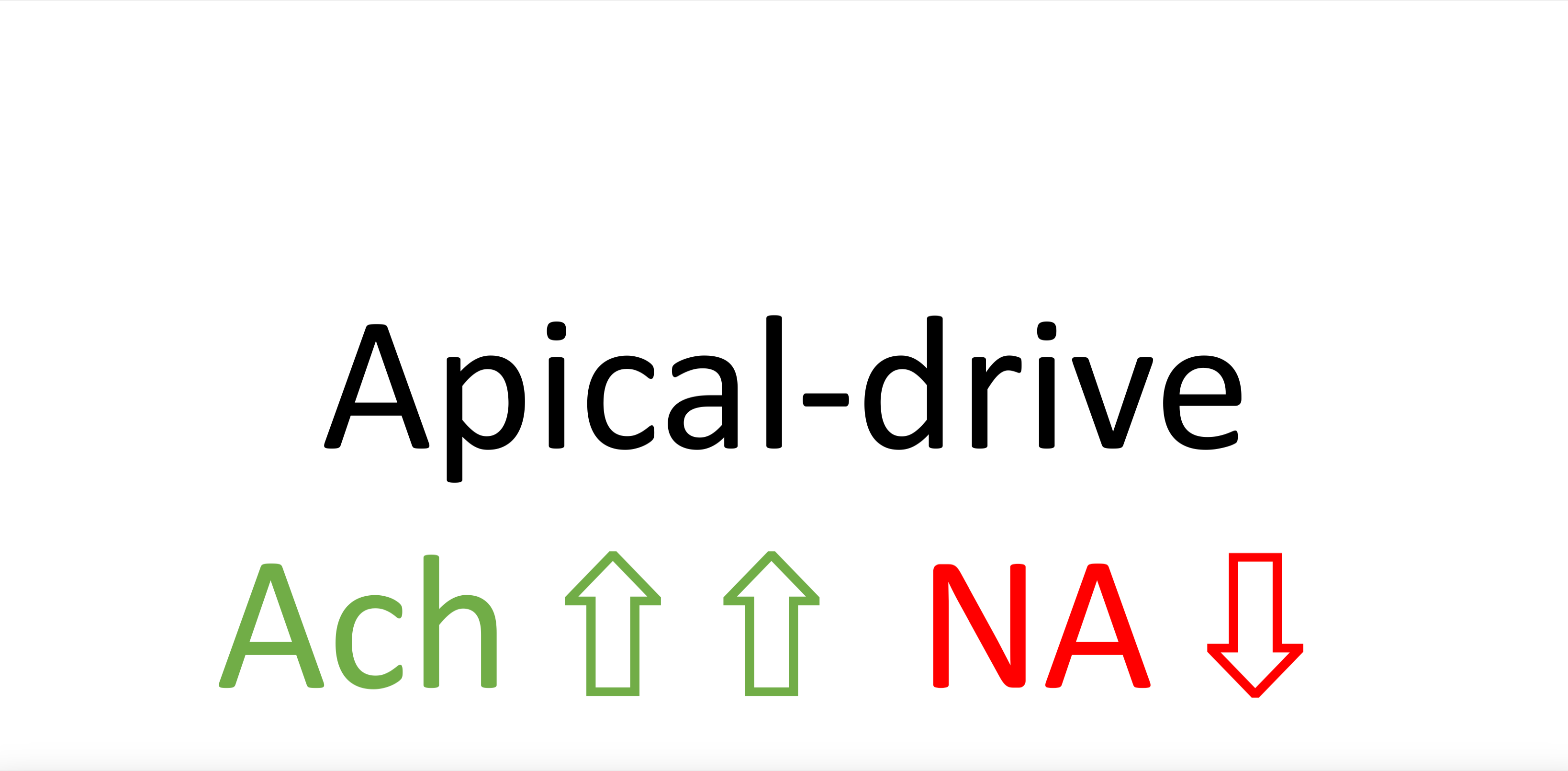}
\end{subfigure}
\begin{subfigure}{.3\textwidth}
  \centering
  \includegraphics[width=0.9\linewidth]{Figures/BrainStates/FR_45_b40_g1_d53_s63.png}
  \caption{}
\end{subfigure}%
\begin{subfigure}{.3\textwidth}
  \centering
  \includegraphics[width=0.9\linewidth]{Figures/BrainStates/FR_45_b200_g03_d58_s68.png}
  \caption{}
\end{subfigure}%
\begin{subfigure}{.3\textwidth}
  \centering
  \includegraphics[width=0.9\linewidth]{Figures/BrainStates/FR_45_b20_g1_d53_s68.png}
  \caption{}
\end{subfigure}
\caption{Apical-amplification, -isolation and -drive: exemplary $\nu(I_s,I_d)$ firing rates produced in the three regimes. Note: max $\nu$ is very different in the tree regimes: over 100Hz in apical-drive, up to 80Hz in -amplification and about 12Hz in -isolation. Also, the jump between the high-firing rate $M_{+}$ and the $M_{-}$ regions spans from tens of Hz in the apical-drive regime down to a few Hz in the -isolation regime.  (a) Apical amplification: $b=40$, $g_C=1$, $E_L^d=-53$, $E_L^s=-63$. (b) Apical isolation: $b=200$, $g_C=0.3$, $E_L^d=-58$, $E_L^s=-68$. (c) Apical-drive: $b=20$, $g_C=1$, $E_L^d=-53$, $E_L^s=-68$. }
\label{fig:Apical-Amplification-Isolation-Drive-3Results}
\end{figure}

Figure \ref{fig:Apical-Amplification-Isolation-Drive-3Results} illustrates the modulation of simulation proxies for ACh and NA to alter the transfer function of the exemplary two-compartment neuron discussed throughout this paper. Specifically, \ref{fig:Apical-Amplification-Isolation-Drive-3Results}.a depicts a representative awake apical-amplification configuration; \ref{fig:Apical-Amplification-Isolation-Drive-3Results}.b presents a configuration tailored to simulate the NREM sleep apical-isolation regime, and \ref{fig:Apical-Amplification-Isolation-Drive-3Results}.c showcases a setting related to the apical-drive configuration, which is expected to be associated with a REM sleep regime.    

%DON'T REMOVE
%Apical-Drive, -Amplification and -Isolation: exemplary results. Plot of the neuron firing rate for somatic and distal input currents in different states. Starting for the apical-amplification configuration prouced by the optimization procedure, Apical drive is obtained (a) Apical drive: the adaptation $b$ is reduced compared to apical amplification, and the soma is hyperpolarized ($b=20$, $g_C=1$, $E_L^d=-53$, $E_L^s=-68$). (b) Apical amplification: all parameters as optimized by L2L framework ($b=40$, $g_C=1$, $E_L^d=-53$, $E_L^s=-63$). (c) Apical isolation, obtained highering the value of the adaptation parameter $b$ with respect of its value in apical amplification, lowering the strength of the coupling between the two compartments and hyperpolarizing both the compartments ($b=200$, $g_C=0.3$, $E_L^d=-58$, $E_L^s=-68$). Note that the range of firing rate in the three plots is very different, reaching values up to 100Hz in apical drive, over 70Hz in amplification and about 12Hz in apical isolation. Analogously, the jump between the two regions with and without calcium activation span from tens of Hz in drive to a few Hz in isolation.

\begin{figure}[ht!]
  \centering
  \includegraphics[width=1.0\textwidth]{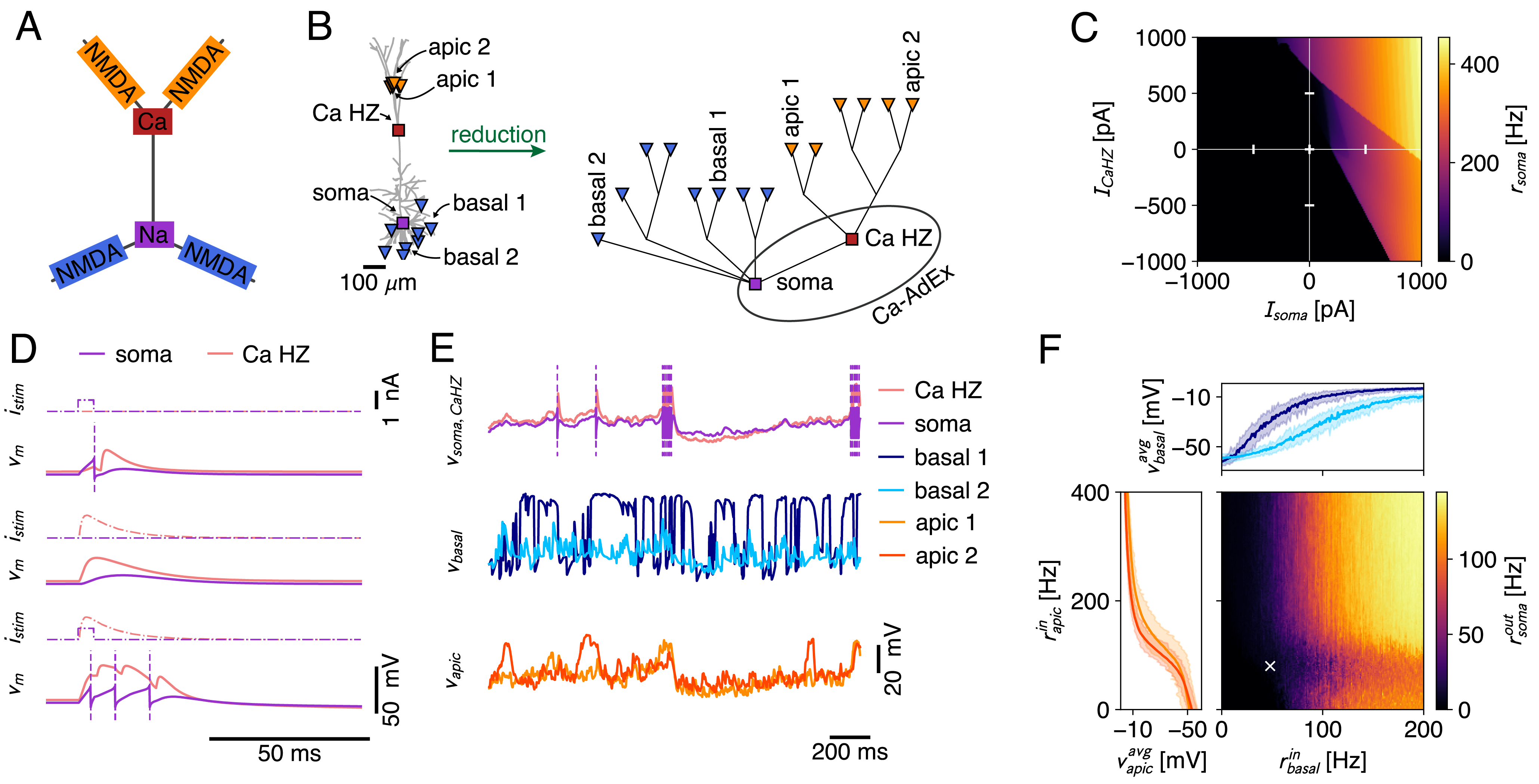}
  \caption{
  \textbf{A:} Canonical view of the interplay between dendritic non-linearities \cite{Larkum2009}. NMDA spikes in distal apical branches (orange) elicit Ca-spikes (red) that result in somatic burst firing (purple), whereas basal NMDA-spikes (blue) directly influence somatic output generation.
  \textbf{B:} Schematic of the creation process of the multicomp AdEx model. A passive morphology (left) with 16 locations (8 basal sites at $\sim$$200$ $\mu$m from the soma [blue triangles], 6 apical tuft sites at $\sim$$1000$ $\mu$m [orange triangles], the Ca-HZ where the apical trunk bifurcates [red square], and the soma [purple square]) is reduced to a simplified compartmental model (right) using the NEAT toolbox \cite{Wybo2021}. The Ca-hotzone and soma are then equipped with the Ca-spike generation mechanism and the AdEx mechanisms, respectively, where parameters were identical to the two compartment model. Labeled apical and basal sites are those for which traces and mean activations are shown in E and F.
  \textbf{C:} Firing rate response to input current steps that were applied to the soma and the Ca-hotzone compartment (same stimulation paradigm as in Figure 2-7).
  \textbf{D:} Simulation of the BAC-firing protocol, where a single output is generated in response to a somatic input pulse (top), no output is generated in response to a Ca-hotzone input (middle), and three output spikes are generated in response to the pairing of inputs (bottom). 
  \textbf{E:} Exemplar traces for stimulation of the model with Poisson inputs that impinged on AMPA+NMDA synapses located at the basal and apical sites. Purple dashed lines indicate spike times.
  \textbf{F:} Firing rate response to increasing input rates to the apical and basal dendritic sites. Axes show the input rate to the individual dendritic sites (input rates were equal across apical resp. basal sites). Inset plots show the average membrane potential in two exemplar apical (left) and basal (top) compartments (same sites as in B, E). The min-max envelope shows the range of values obtained over all activation levels of the other area (i.e. apical vs basal). The white cross marks the input rates shown in E. 
  }
  \label{fig:nmda}
\end{figure}

\subsection{Extending the two-compartment layout}

While important, the Ca$^{2+}$-spike is not the only dendritic event that fundamentally shapes the neuronal input/output relation (Figure \ref{fig:nmda}A). 
Through the voltage dependent unblocking of the NMDA-receptor channel \cite{MacDonald1982, Jahr1990a}, coincident inputs to dendritic branches summate supra-linearly, and the resulting events are known as NMDA-spikes \cite{Schiller2000, Major2008, Major2013}. 
As our Ca-AdEx framework is embedded in a general compartmental modelling framework, it is straightforwardly possible to extend the two-compartment description to one where there are additional dendritic subunits that can produce NMDA-spikes. 
We demonstrate the potential of our approach by deriving the parameters of these subunits from a realistic L5PC morphology (Figure \ref{fig:nmda}B, \cite{Hay2011}), using the method based on resistance matrix fits proposed by Wybo et al. \cite{Wybo2021}. 
The somatic and Ca-HZ compartment are then respectively equipped with the AdEx and Ca$^{2+}$-spike mechanisms, where we used the same parameters as the two-compartment model (Table \ref{tab:GenomeTable}).
It speaks to the robustness of our approach that we achieve qualitatively similar behaviour as the two-compartmental model, without refitting any of the parameters (Figure~\ref{fig:nmda}C). 
Furthermore, this extended Ca-AdEx model also reproduced the BAC-firing protocol (Figure~\ref{fig:nmda}D).
We then equip the apical (Figure \ref{fig:nmda}B, orange) and basal (Figure \ref{fig:nmda}B, blue) compartments with excitatory synapses containing both AMPA and NMDA receptor channels, as well as with an inhibitory GABAergic synapse.
The latter was stimulated with a fixed Poisson rate of 20 Hz, whereas for the former we scanned a range of firing rates: for the apical synapses, we delivered Poisson rates between 0 and 400 Hz in 2 Hz increments, while for the basal synapses Poisson rates between 0 and 200 Hz were probed, in 1 Hz increments (Figure \ref{fig:nmda}E,F). 
Simulations with each set of input rates were run for 2000 ms, and the average output rate was measured by averaging over five such episodes.
The dendritic voltage traces exhibit signatures of the nonlinear dynamics associated with NMDA- and Ca$^{2+}$-channels (i.e. long up-states, burst firing, etc; Figure \ref{fig:nmda}E). 
Furthermore, the averaged voltage responses in the apical and basal subunits follow the typical sigmoidal response curve \cite{Schiller2000, Major2008, Branco2010, Poirazi2003, Singh2015} (Figure \ref{fig:nmda}F, insets).
These inset plots show the min-max envelope of the averaged voltage, i.e. for the apical voltage response (orange), the minimal values occurs for the lowest basal input level, whereas the maximal value occurs for the highest basal input level.
That these min-max envelopes are close together and do not substantially affect the sigmoidal response curve, demonstrates that apical and basal areas are mutually independent \cite{Wybo2019}.
Finally, the supralinear Ca$^{2+}$-spike mediated interaction between apical and basal areas is clearly visible in the output firing rates, where a strong increase occurs above a 100 Hz basal and a 100 Hz apical input rate (Figure \ref{fig:nmda}F).
We also remark that while the input and output rates seem high when considered as tonic firing rates, it is reasonable to assume that such rates can and do occur transiently, through the coincidence of multiple inputs to the apical and/or basal regions.

\section{Discussion}
\label{sec:Discussion}
%Support of brains state specific apical-amplification, -isolation and -drive
Here we present the Ca-AdEx model, capturing the essential features of the apical-amplification, -isolation, and -drive regimes, at a modest computational cost compared to classical point-like neurons. This advancement supports the development of network models capable of emulating awake, NREM, and REM-like states, as well as the learning capabilities associated with the emergence of brain-state-specific bursting regimes in neurons that detect the coincidence of apical and somatic signals.
Apical mechanisms play a crucial role in optimally combining internal priors and perceptual evidence within multi-areal hierarchical systems featuring lateral, top-down, and bottom-up connections.
A significant observation is the drastic change in the firing rate of neurons where apical-amplification is active, which facilitates learning that aligns with the higher sampling rate of world experiences and is compatible with the STDP window of a few tens of milliseconds. Notably, an even more enhanced firing regime is associated with the apical-drive condition, potentially related to the replay and association of experiences during dreaming, whereas the suppression of the effect in apical-isolation supports the loss of consciousness during deeper sleep stages.
%Created a simplified transfer function for non-spiking models generalizing RELUs (PSP)
Furthermore, as detailed in section \ref{subsec:TransfFuncDesc}, it is feasible to formulate, at a high level of abstraction, a compact geometric model capturing the effects produced by the combination of signals that convey information about priors and perceptual evidence, segregated into the apical and somatic compartments.
The transfer function of the two-compartment Ca-AdEx model described here can be approximated piece-wise by low-order polynomials. Specifically, we examined the case of two approximating planes, giving rise to a class of transfer functions named ThetaPlanes($I_s,I_d$). ThetaPlanes represents a generalization for two-compartment neurons of the ReLU function commonly used to approximate single-compartment neuron models in numerous artificial intelligence algorithms. 
%Future: Relation with artificial intelligence literature (PSP)
ThetaPlanes transfer functions can be implemented as efficient computational gates for use in large cognitive networks at a high level of abstraction. In future works, we plan to investigate the benefits of this computational gate in next-generation bio-inspired artificial intelligence algorithms. This expectation is supported by the emerging value of brain-state-specific bursting regimes demonstrated in recent works \cite{CaponeLupoMuratorePaolucci2023, CaponeLupoMuratorePaolucci2022}. These studies, while assuming the existence of such coincidence detection mechanisms as working hypotheses, lacked a biologically grounded transfer function.

%relation with TVB and whole-brain models
Furthermore, the potential to maintain compatibility with the transfer function of widely adopted leaky integrate-and-fire models with adaptation when apical amplification is not triggered is promising. In our case, we aimed for compatibility with the Adaptive Exponential Integrate-and-Fire model (AdEx), which is extensively used for simulations at both micro- and meso-scales. It also serves as the basis for mean-field models for simulations encompassing the whole cortex \cite{CaponeDeLucaDeBonisKetaMouse2023}.
We anticipate that during wakefulness, apical mechanisms and the sparsity of long-range connections will place a strict minority of neurons in a bursting regime. This adjustment is unlikely to significantly alter the average spectral signatures of expressed rhythms but could induce profound effects on perception and learning ability. Such a balance is necessary to maintain compatibility with the extensive body of experimental evidence concerning rhythms, average firing rates, and their fluctuations.
During sleep, we anticipate that a delicate balance will be maintained to ensure healthy sleep patterns and to promote its beneficial cognitive and energetic effects.
% identification of candidate neurons thanks to L2L 
An additional noteworthy observation is that two-compartment neurons with significant transfer functions were efficiently discovered using the L2L framework within an evolutionary process that spanned only a hundred generations, each including no more than a hundred individuals.
%Natural evolution can easily discover the advantages of apical mechanism using a simple 2C geometry
In our view, this suggests that natural evolution could have readily identified the cognitive advantages of apical mechanisms through localized variations of membrane and channel parameters, in ways somewhat analogous to the creation of two compartments. Thus, evolution might have incrementally given rise to the complex morphology seen in pyramidal neurons in the cortex.

Another aspect touched by our work is the role of high-performance computing (HPC) infrastructure, which offers a platform for conducting increasingly robust, comprehensive, and extensive explorations of parameter spaces in scientific models. Coupled with machine learning, HPC emerges as a potent digital environment for adaptive testing and understanding the interactions between data and models. HPC allows scientists to simultaneously test a vast number of hypotheses within short time frames, delivering crucial information that can be incorporated into accelerated experimental cycles. Within this framework, L2L serves as an accessible tool for domain scientists to interface with HPC and conduct efficient parameter explorations. It allows focusing on areas of interest while offering a comprehensive overview of the entire parameter space, including the relationships between parameters and the selected fitness metrics. In this manuscript, we demonstrated that L2L is a framework adept at leveraging HPC infrastructure to assist neuroscientists in optimizing, fitting, and searching for suitable dynamics in models. Specifically, following the definition of the genome and the fitness functions for the multi-compartment neuron, an evolutionary algorithm can identify suitable candidates that survive the selection process.
This work, based on a customization of the multi-compartment framework available in NEST \cite{Gewaltig2007, Spreizer2022}, also facilitates the inclusion of two- and many-compartments neuron models supporting apical mechanisms in the ecosystem of other standard simulation engines like Neuron \cite{Carnevale2006} and Brian \cite{Stimberg2019}.
Additionally, this work outlines an approach grounded in traditional compartmental dynamics, which is computationally efficient and accurately captures the interplay between somatic action potentials (APs) and dendritic \Ca-spikes. As part of a broader compartmental modeling framework in NEST, our model can easily be expanded with additional compartments to represent other dendritic events, such as N-methyl-D-Aspartate (NMDA) spikes.
Finally, due to its implementation in NEST, the model can be directly integrated into network simulations modeling incremental learning and sleep cycles.

%Future: Insertion in a standard simulation engine 

\section{Source code}
\label{sec:SourceCode}
\pier{To be released on paper submission or by direct contact to start research partnership.}

\section{Exemplary two-compartment Ca-AdEx neuron parameters}
\label{sec:ExemplaryParams}
The evolutionary search detailed in the \nameref{sec:Methods} Section identified an exemplary individual utilized throughout this paper unless specified otherwise (for instance, when discussing modulation to other brain states). Its complete genome is provided below.
The parameters of the corresponding \textit{ThetaPlanes} fitting function, as discussed in Section \ref{subsec:TransfFuncFit}, are also reported below.
\pier{To be released on paper submission or by direct contact to start research partnership.}

\section{Acknowledgments}
This work has been co-funded by the European Next Generation EU grants CUP I53C22001400006 (FAIR PE0000013 PNRR Project) and CUP B51E22000150006 (EBRAINS-Italy IR00011 PNRR Project) and by the European Union Horizon 2020 Research and Innovation program under the FET Flagship Human Brain Project (grant agreement SGA3 n. 945539). Also, we acknowledge the usage of FENIX infrastructure computational resources under the ICEI project (grant agrement no. 800858) attributed to Chiara De Luca. In the end, we acknowledge the support of the APE Parallel/Distributed Computing Laboratory of INFN, Sezione di Roma.
This research has also been partially funded by the Helmholtz Association through the Helmholtz Portfolio Theme Supercomputing and Modeling for the Human Brain. 

\section{Individual Contributions}
\begin{itemize}
\item E.P. and P.S.P. Scientific conception, definition of neural activity equations and evolutionary fitness functions, definition of ThetaPlanes funtions, analysis of results, initial manuscript writing.
\item A.Y. and S.D. Learning to learn tool for evolutionary selection of optimal neural parameters, initial manuscript writing, support on HPC platforms.
\item N.K. Fitting of ThetaPlanes functions, analysis of results, initial manuscript writing.
\item W.W. Framework for multi-compartment neuron modeling in NEST, integration of Ca-AdEx in extended compartmental models, initial manuscript writing.
\item F.S. Computational resources and system support.
\item J.F.S. Definition of neural activity equations, biological plausibility of the multi-compartment neuron, initial manuscript writing
\end{itemize}

\printbibliography

\end{document}